\documentclass[14pt]{extarticle}

\usepackage{amssymb,amsmath,amscd}
\usepackage{bm}
\usepackage{subfigure}
\usepackage{multicol}

\usepackage{mathtext}         
\usepackage[makeroom]{cancel} 
\usepackage[utf8]{inputenc}  

\usepackage{amssymb}                                          
\usepackage{amsmath}                                          
\usepackage{amssymb}

\usepackage{graphics}
\usepackage{graphicx}
\usepackage{changes}

\usepackage{xcolor}
\usepackage{wrapfig,epsfig,epsf}
\usepackage{bm}    
\usepackage{hyperref}

\makeatletter
\renewcommand\@biblabel[1]{#1.} 
\makeatother

\def\apjl{ApJL}
\def\apjs{ApJ Suppl. Ser.}

\def\aap{A\&A}

\usepackage[left=2.5cm,right=2.5cm,top=4cm,bottom=2.7cm]{geometry}

\def\beq#1{\begin{equation}\label{#1}}
\def\eeq{\end{equation}}
\def\beqa#1{\begin{eqnarray}\label{#1}}
\def\eeqa{\end{eqnarray}}

\def\myfrac#1#2{\left(\frac{#1}{#2}\right)}

\def\comment#1{\relax}

\def\dfrac#1#2{\displaystyle\frac{\partial #1}{\partial #2}}

\title{Pulsar timing and polarimetry: results and perspectives}
\author{
K.~A.~Postnov$^{1,2}$,\\
N.~K.~Porayko$^{3,4}$,\\
M.~S.~Pshirkov$^{1,5,6}$\\
$^1$ Sternberg Astronomical Institute, \\Lomonosov Moscow State University, Universitetsky prospekt 13, \\119992, Moscow, Russia\\
 $^2$ Kazan Federal University, \\ Kremlyovskaya str 18, Kazan, 420008
 Russia\\
$^3$ Max-Planck-Institut f{\"u}r Radioastronomie,\\ Auf dem H{\"u}gel 69, 53121 Bonn, Germany\\
$^4$Dipartimento di Fisica “G. Occhialini”,\\ Universit{\"a} degli Studi di Milano-Bicocca,\\ Piazza della Scienza 3, I-20126 Milano, Italy 
}
\begin{document}
\date{}
\maketitle
\tableofcontents
\begin{abstract}
Pulsar timing, i.e. the analysis of the arrival times of pulses from a pulsar, is a powerful tool in modern astrophysics.
It allows us to measure the time delays of an electromagnetic signal caused by a number of physical processes as the signal propagates from the source to the observer. Joint analysis of an ensemble of pulsars (Pulsar Timing Arrays, PTAs) can be used to address a variety of astrophysical challenges, including the problem of direct detection of space-time metric perturbations, in particular those induced by gravitational waves.
Here we present a comprehensive review of the current state of research in the field of pulsar timing, with particular emphasis on recent advancements in the detection of stochastic background of nHz gravitational waves, reported by a number of international collaborations such as NANOGrav (North American Nanohertz Observatory for Gravitational Waves), European Pulsar Timing Array (EPTA), Chinese Pulsar Timing Array (CPTA) and Indian Pulsar Timing Array (InPTA), which are joining their efforts within the International PTA (IPTA). Additionally, this paper reviews contemporary constraints on scalar ultralight matter (pseudoscalar bosons), obtained from timing and polarimetry data of pulsars. The prospects for applying these tools to other problems in fundamental physics and cosmology are explored.
\end{abstract}

Keywords: radio pulsars, 
pulsar timing, gravitational waves, stochastic background of gravitational waves, supermassive double black holes, ultralight scalar dark matter

\everymath{\displaystyle}
\section{Introduction}
\label{intro}

In this paper, we review the method of pulsar timing (PT), which is a technique for the precise measurement of the times of arrival (ToAs) of pulses from pulsars acting as stable natural cosmic clocks.
Pulsar timing is an essential tool in the search for gravitational waves (GWs) both from individual sources and the stochastic GW background (GWB). Two seminal 
theoretical works by V.A. Rubakov are of particular relevance in this area of modern experimental astrophysics. He was among the first to calculate the stochastic GWB from the inflationary stage of the expansion of the Universe \cite{1982PhLB..115..189R}, which is one of the viable targets of the PT experiment. In addition, he proposed a methodology of searching for a monochromatic signal from ultralight scalar dark matter (DM) in the Galaxy using PT data \cite{2014JCAP...02..019K}.

The search for low-frequency GWs using PT is a concept of detecting distortions of the space-time metric by precisely measuring the ToAs of signals from the highly stable periodic sources (natural or artificial "clocks"). 
The application of pulsar timing for low-frequency GWs was proposed in the pioneering works of M.V. Sazhin \cite{1978SvA....22...36S} and S. Detweiler \cite{1979ApJ...234.1100D}\footnote{See also the earlier work \cite{1975GReGr...6..439E}, where the authors calculated the GW-induced variation of a frequency of a periodic signal received on Earth during Doppler tracking of a spacecraft.}. Indeed, in the simplest case, the trajectory of a photon with a frequency $f^\textrm{em}$ in the field of a monochromatic plane-polarized 
GW and amplitude $h_+\ll 1$ (for simplicity only "+" polarization is considered) propagating along the $z$-axis, which is perpendicular to the line of sight (LoS)  to the source, residing at distance $D$, is described by the equation $ds^2=0=-c^2dt^2+(1+h_+(z,t))dx^2+(1-h_+(z,t))dy^2+dz^2$.

Travel time of such a signal from the source to the observer along the $x$ axis will change by the amount 
$$
\delta t=\frac{D}{c}[(1\pm\frac{1}{2}h_+(t))-(1\pm \frac{1}{2}h_+(t-D/c)] =\pm \frac{D}{2c}(h_+(t)-h_+(t-D/c)),
$$ which corresponds to the shift in the frequency of the received signal \footnote{The expression for the redshift of the observed frequency in the GW field can be derived in a straightforward way from the Sachs-Wolfe formula, originally used to describe the fluctuations in the temperature maps of the cosmic microwave background (CMB): \cite{1967ApJ...147...73S,1985SvAL...11..133S}: $\frac{\Delta T}{T}=-\frac{1}{2}\int_{\lambda_e}^{\lambda_o}\dfrac{ h_{ij}}{ t}\hat n_i\hat n_jd\lambda$ ($\hat n_i, \hat n_j$ is unit vector in the pulsar direction; the integral is taken along the photon path)
}:  
$\Delta f^\textrm{em}(t)/f^\textrm{em}=\frac{1}{2}(h_+(t)-h_+(t-D/c))$. The delays in a ToAs of periodic pulses accumulated during the observation time $T$ are $\mathcal{R}(T)=\int_0^T \frac{\Delta f^\textrm{em}(t)}{f^\textrm{em}} dt$.

In the general case, a plane GW with frequency $f=\omega/2\pi$, dimensionless amplitude $h_c$ and polarization angle $\phi$ propagating at an angle $\theta$ to the LoS of a pulsar, will cause sinusoidal oscillations in the ToAs of its pulses with characteristic amplitude \cite{2004ApJ...606..799J,2011ApJ...730...29W}
\begin{equation}
\label{e:TOA_GW}
\Delta t=\frac{h_c}{\omega}[1+\cos\theta]\sin 2\phi\sin{\frac{\omega D[1-\cos\theta]}{2c}}\,,
\end{equation}
where $c$ is the speed of light (below, unless specified, we use the natural system of units $\hbar=c=k_B=1$).

The largest magnitude of the effect is achieved for a total observing time of the order of $T_\textrm{obs}=1/f$. For a characteristic timespan of observations of several years ($T_\textrm{obs}\sim 10^8$~s), the amplitude of the ToA variations will be of the order of $\mathcal{A}=h_c/\omega\sim 20$ ns for $h_c\sim 10^{-15}$. After subtracting the known deterministic contributions encapsulated in the pulsar timing model due to the motion of the observatory in the gravitational field of the Sun and planets in the Solar System and taking into account the orbital motion of a pulsar in a binary system, as well as removing the effects of General Relativity (GR) (R{\"o}mer delay, gravitational redshift, Shapiro delay), \textit{cleaned} ToA residuals (also known as \textit{post-fit} residuals) can be obtained. Analysis of post-fit residuals with typical root-mean-square (RMS) of several tens of ns can be used to find various effects related to fluctuations of the space-time metric, including GWs from individual supermassive black hole binaries (SMBHBs), stochastic GWB, both of astrophysical and cosmological nature, as well as some forms of DM (in particular, ultralight scalar DM causing oscillations of the gravitational potential considered for the first time by A. Khmelnitsky and V. Rubakov \cite{2014JCAP...02..019K}). A detailed review of the astrophysical and cosmological sources of nHz GWs to which the PT method is most sensitive can be found in \cite{2019A&ARv..27....5B}

This review focuses on these and other possible sources of space-time nonstationarities that are currently studied with PT.
The structure of the review is as follows. Section~\ref{s:PT} briefly presents the essence of the PT method. Section \ref{s:GWSB} discusses the stochastic GWB and summarize the results on its detection and interpretation by international PT collaborations. Section~\ref{s:ULDM} is devoted to probes of possible traces of ultralight scalar DM in PT (subsection \ref{s:uldm_SW}) and pulsar polarimetry (subsection \ref{s:uldm_birefr} ). Section~\ref{s:other} discusses other possible applications of PT, i.e. detection of GW bursts with memory, searches for additional modes of GW polarization, constraints on the graviton mass and GW propagation velocity, study of the mass distribution in the Galaxy, and search for compact objects.
In conclusion (Section~\ref{s:conclusion}), the main trends in the advancement of the PT method in solving astrophysical and cosmological problems are briefly described.

\section{High-precision pulsar timing} 
\label{s:PT}
\subsection{Brief description of the method}
PT can undoubtedly be called a unique instrument for detecting GWs in the nHz frequency range. Pulsars are rapidly rotating strongly magnetized neutron stars. In the PT context, these cosmic flywheels act as ultra-stable clocks located at Galactic distances. But prior to using pulsars as stable clocks, it is necessary to perform a number of reductions, which we describe below.

Individual pulses from pulsars vary considerably in shape and amplitude, but their mean profiles, obtained by averaging a large number (in the case of millisecond pulsars (MSPs) we are talking about $~\mathcal{O}(10^6)$) of individual pulses, are extremely temporally stable and are a kind of "business cards"\  of pulsars. 
During a single observing session, the signal is accumulated for a relatively long time (from minutes to tens of minutes), then the signal is \textit{folded} with a period equal to the spin period of the pulsar at that epoch. The resultant folded profile is then cross-correlated with the mean \textit{template} profile. The outcome of an individual observational session is a ToA, which is typically referred to the time stamp closest to the middle of the session. Accuracy of the measured ToA is approximately determined by ${W}/{\mathrm{SNR}}$, where $W$ is the pulse width (not to be confused with the pulsar period $P$) and $\mathrm{SNR}$ is the signal-to-noise ratio (SNR) of the observed profile \cite{2012hpa..book.....L}:
\begin{equation}
\mathrm{SNR} \propto S_{\mathrm{psr}}\sqrt{t_{\mathrm{obs}}\Delta f}. \sqrt{\frac{P}{W}},
\label{eq:timing_SNR}
\end{equation}

where $S_\mathrm{psr}$ is the mean flux density of the pulsar, $t_{\mathrm{obs}}$ is the total duration of the session, and $\Delta f$ is the bandwidth of the instrument. It is clear that SNR is larger for bright pulsars with sharp profile shapes (so that the $W/P$ ratio is smaller) observed over a longer period of time with a wider frequency bandwidth. In absolute values, the RMS error of the determined ToA $\sigma_n=\sqrt{\langle\mathcal{R}^2(t)\rangle}$ is smaller for MSPs, which, together with their high rotational stability, explains their usefulness in the search for GW and other physical effects that is the subject of this review.

Due to pulsar's gradual loss of rotational energy, its spin frequency $\nu$ slowly decreases \cite{Beskin:2018}. This slow-down can be described as follows:
\begin{equation}
    \nu (t)=\nu (t_0)+\dot{\nu }_0(t-t_0)+\frac{1}{2}\ddot{\nu }{_0}(t-t_0)^2+\ldots,
    \label{eq:freq_evolution}
\end{equation}
hereafter, the index 0 denotes the value at time $t_0$, e.g., $\nu _0 \equiv \nu (t_0)$.

One can easily obtain the value of the pulsar phase at any given time $t$ with respect to the initial moment $t_0$:
\begin{equation}
N(t)=N_0+\int_{t_0}^t \nu(t)dt=N_0+\nu _0(t-t_0)+\frac{1}{2}\dot{\nu }_0(t-t_0)^2+\frac{1}{6}\ddot{\nu }_0(t-t_0)^3 + \ldots.
\label{eq:num_pulse}
\end{equation}
This relation holds true in the pulsar reference frame. To convert it to the reference frame of a ground-based observatory, appropriate reduction must be made.

For the relation (\ref{eq:num_pulse}) to remain valid, it is necessary to use a time scale associated with a reference frame that is as close as possible to the inertial one. At present, such a reference frame is a barycentric coordinate system, with the origin in the barycenter of the Solar system, and the corresponding Barycentric Coordinate Time (TCB, Temps-Coordonn{\'e}e Barycentrique).

We will consider an isolated single pulsar and assume that it is at rest with respect to the chosen reference frame. A uniform motion will not change the form of Eq.~(\ref{eq:num_pulse}), but will only redefine the observed value of the spin frequency of a pulsar compared to its intrinsic value, due to the Doppler shift.

First, ToAs at the observatory are calculated  w.r.t a local timescale defined by the local hydrogen standard. Further reduction is carried out in a series of steps. At the first stage, using global navigation satellite systems such as GPS or GLONASS, the local time scale is connected to the Coordinated Universal Time (UTC), rigidly linked to the International Atomic Time (TAI) scale, which is a weighted average of several hundreds of atomic clocks at various national laboratories. The next step is a reduction to the most stable Terrestrial Time (TT) scale, which is the closest possible approximation to an ideal uniform atomic clock at the level of the local geoid.

Finally, the ToAs of the pulse received at the telescope are recalculated to the barycenter of the Solar System. In this process, high-precision ephemerids are used, such as those developed by NASA Jet Propulsion Laboratory (JPL NASA) DE440, or IPA RAS EPM2021. In addition to the R\"omer delay between the telescope and the Solar system barycenter, relativistic effects, such as Shapiro delay and the effect of time dilation on the moving Earth and in the gravitational fields of Solar system bodies, are taken into account \cite{1986ARA&A..24..537B,1990SvA....34..496D}. Currently, the accuracy with which we can account for all of these effects is about 100~ns, which is equivalent to knowing the position of the barycenter with a precision up to tens of meters. Even such small uncertainties can bias the detection of GWs with PT \cite{2020ApJ...893..112V}.

The ToA obtained in the TCB is the main result of this reduction. This procedure of ToA extraction is carried out over many years with a cadence of several weeks, forming a long time series of ToAs. The most up-to-date PT ephemeris are obtained using the least-squares fit of the PT model to this ToA time series:

\begin{equation}
\chi^2=\sum_i \left(\frac{N(t_i)-n_i}{\sigma_n^i}\right)^2,
\label{eq:chi2_timing}
\end{equation}
where $N(t_i)$ is the pulse number corresponding to the ToA, $t_i$, $n_i$ is the integer number closest to $N(t_i)$, since for an ideal model between each pair of ToAs, an integer number of rotations of the pulsar must have occurred; $\sigma_n^i$ is the measurement error of the $i$-th ToA in units of pulsar period.
Given that there are a number of effects that are not accounted for by the pulsar timing model, $N(t_i)\neq n_i$  and ToA residuals are non-zero: $\mathcal{R} \equiv P(N(t_i)-n_i)$.
In practice, this least-square fit is routinely performed using TEMPO\footnote{\url{https://tempo.sourceforge.net/}} and TEMPO2 software packages \cite{2006MNRAS.369..655H,2006MNRAS.372.1549E}.

If the PT model ephemeris have been correctly determined, i.e., when they are close to their true values, the residuals behave like  "white noise"\  with RMS close to the ToA uncertainty in individual sessions, $\sigma_n^i$. If there is an error in the parameter values, the residuals display a characteristic structure that enables one to refine the ephemeris. For example, sinusoidal oscillations with an annual period would indicate incorrect determination of the celestial coordinates of the pulsar that lead to a wrong estimation of the R\"omer delay and errors in the reduction to the barycenter of the Solar system. That is, PT is an iterative tool for fitting and refining of the PT model parameters when new data are added at each consecutive step. 

The presence of a GWB will give rise to a "red noise"\footnote{The power spectral density increases with decreasing frequency.} in the residuals. One should note that residuals with similar spectral properties can also be induced by other astrophysical processes, such as irregularities in pulsar rotation (pulsar intrinsic "red noise"), noise in the atomic time standards which are used to measure ToAs at the observatory, and turbulence of the interstellar medium\footnote{For instance, the latter causes a stochastic signal in the residuals with the amplitude depending on the observing frequency.}. 
The main difficulty in the GWB detection using the PT method lies in separating the sought signal from other aforementioned stochastic contributions. The non-trivial spatial correlation of the GW signal, which will be discussed in the next subsection, greatly simplifies this task.

\subsection{Correlations of ToAs from different pulsars}
\label{sec:h-d}

The quadrupole nature of GWs leads to a unique correlation in the ToAs of pulsars that depends on the angular separation between the pulsars in pairs.  If there is an isotropic stochastic GWB with equal power in all directions, the correlation of the ToA residuals from a pair of pulsars with an angular separation $\gamma_{ij}$ between them is described by the so-called Hellings-Downs (HD) curve \cite{1983ApJ...265L..39H}:
\beq{e:HellinsDowns}
    \alpha_{ij}=\frac{1}{4\pi}\int\alpha_i\alpha_jd\Omega=\frac{1-\cos\gamma_{ij}}{2}\ln\myfrac{1-\cos\gamma_{ij}}{2}-\frac{1}{6}\frac{1-\cos\gamma_{ij}}{2}+\frac{1}{3}
\eeq
(see \cite{2009PhRvD..79h4030A, Bernardo:2024bdc} for the detailed derivation and discussion of the limitations of this formula).

The effect of GWs on pulsar ToAs includes two contributions, the "Earth term"\ and  the "pulsar term"\ , which correspond to the metric perturbation at Earth and at the pulsar, respectively. Since distances to pulsars are poorly known, it is usually possible to correlate only the common "Earth"\ part. This leads to a "blurring" of the HD dependence (known as \textit{HD variance}). Adding pulsar distances as free parameters in the analysis can lead to an increase in the SNR of the GW signal from an individual SMBHB and a more accurate estimation of the distance to this source \cite{2010arXiv1008.1782C}. For a stochastic GWB, as shown in \cite{2023PhRvD.108d3026A}, the \textit{pulsar variance}\ of the HD curve can be suppressed by increasing the number of observed pulsars and then averaging over multiple pulsar pairs over some range of angular distances $\gamma_{ij}$.

Another effect that leads to scattering of the expected HD curve is related to the so-called \textit{cosmological variance}\
\cite{2022JCAP...11..046B, 2023PhRvD.107d3018A, 2023PhRvD.108d3026A}. Cosmological variance is caused by the interference of several GW sources emitting at the same frequency, which leads to a deviation from the canonical curve. The specific form of the modified HD curve will depend on the realization of an ensemble of SMBHBs in the observed Universe. Unlike the pulsar variance, the cosmological variance cannot be suppressed by increasing the number of pulsars in the analysis.

\section{Effect of a stochastic GWB on PT}
\label{s:GWSB}

A stochastic GWB is generated by a population of independent sources of GWs, and is treated as a random variable over an ensemble of realizations. It is conveniently described by the dimensionless spectral amplitude $h_c(k,t)$, which for a homogeneous and isotropic unpolarized random Gaussian background is:
\begin{equation}
    \langle h_l(\mathbf{k},t) h_m(\mathbf{q},t)\rangle = \frac{8\pi^5}{k^3}\delta^3(\mathbf{k}-\mathbf{q})\delta_{lm}h_c^2(k,t)
\end{equation}
where $k=|\mathbf{k}|$. 
Given this definition, the relation with the random tensor amplitudes of the metrics of two polarizations is given by the expression 
\begin{equation}
    \langle h_{ij}(\mathbf{x},t) h_{ij}(\mathbf{x},t)\rangle=2\int_0^\infty\frac{dk}{k}h_c^2(k,t)\,.
\end{equation}

It can be shown that $h_c(k,t)$ is the characteristic amplitude of the metric in the logarithmic interval of wavenumbers per one polarization. A common practice is to introduce the power spectrum $P_h(k,t)\equiv 2h_c^2(k,t)$. 
The energy density of the GWB is
\begin{equation}
    \rho_{GW}=\int\frac{dk}{k}\frac{d\rho_{GW}}{d\log k}=\frac{ \langle \dot{h}_{ij}(\mathbf{x},t) \dot{h}_{ij}(\mathbf{x},t)\rangle}{32\pi G}\,.
\end{equation}
To relate to observed quantities, one can introduce a one-sided GWB spectral density $S_h(f) [1/\hbox{Hz}]$ in terms of frequency $f=k/(2\pi)$
\begin{equation}
    S_h(f)=\frac{h_c^2(f)}{2f}
\end{equation}
which is convenient to relate to the noise power spectral density of a receiver $S_n(f)$. 
GWB (especially of cosmological nature) is conventionally characterized in terms of the dimensionless ratio of the GW energy density per logarithmic frequency interval to the critical density of the Universe $\rho_c=3H_0^2/(8\pi G)$ (where $H_0$ is the present value of the Hubble parameter):
\begin{equation}
\label{e:Omegagw}
    \Omega_\mathrm{GW}(f)=\frac{4\pi^2}{3H_0^2}f^3S_h(f)=\frac{2\pi^2}{3H_0^2}f^2 h_c^2(f)
\end{equation}
In terms of the dimensionless amplitude of metric $h_c=\sqrt{2fS_h}$
\begin{equation}
    S_h(f)\approx 8\times 10^{-37}[\hbox{Hz}^{-1}]\myfrac{\hbox{Hz}}{f}^3\myfrac{H_0}{100\hbox{km/s/Mpc}}^2\Omega_\mathrm{GW}(f)
\end{equation}
\begin{equation}
    h_c\approx 1.26\times 10^{-18}\myfrac{\hbox{Hz}}{f} 
    \myfrac{H_0}{100\hbox{km/s/Mpc}}\sqrt{\Omega_\mathrm{GW}(f)}\,.
\end{equation}

When searching for stochastic GWB with PT, we use the power spectral density of the  correlated residuals \cite{2013PhRvD..88l4032T}. In this case, instead of the GWB spectral density $S_h(f)$ we use the one-sided spectral density of the residuals $S_n(f)$ with dimensionality 1/Hz$^3$ 
\footnote{The extra square of frequency in the denominator of the residuals' power spectrum appears because the RMS of the residuals  is  expressed  as $\langle \mathcal{R}^2(t) \rangle=2\int_0^\infty\frac{1}{3} \frac{h_c^2(k)}{k^2}\frac{dk}{k}=2\int_0^\infty df S_n(f)$ (the factor 1/3 arises when averaging over the distances to the pulsars under the condition $\omega D\gg 1$). Since the residuals are in [s], the power spectrum $S_n(f)$ has the dimension [s$^2$/Hz] = [Hz$^{-3}$].}
\begin{equation}
\label{e:Sfhc}
     S_n(f)=\frac{1}{12\pi^2}\frac{h_c^2(f)}{f^3}\,.
\end{equation}
From Eq.~(\ref{e:Omegagw}) we obtain
\begin{equation}
\label{e:Sfomegagw}
     S_n(f)=\frac{H_0^2}{8\pi^4}\frac{\Omega_\textrm{GW}(f)}{f^5}\,.
\end{equation}

The RMS of the induced residuals $\sigma_{\mathrm{ToA}}$ with a total time span $T_\textrm{obs}$ in the frequency bins $\Delta f_i=1/T_\textrm{obs}$ are related to the spectral density $S_n(f)$:
\begin{equation}
\sigma_{\mathrm{ToA},i}=\left(\int_{\Delta f_i}S_n(f)df\right)^{1/2} \approx \left(S_n(f_i)\Delta f_i\right)^{1/2}=\myfrac{S_n(f_i)}{T_\textrm{obs}}^{1/2}\,.
\end{equation}
For characteristic values $T_\textrm{obs}=1$~year and $h=10^{-15}$ we obtain:
\begin{eqnarray}
\label{e:sigmaTOA}
    &\sigma_\mathrm{TOA}\approx 1.6\times 10^{-8} [\hbox{s}]\myfrac{h_c}{10^{-15}}\myfrac{f}{10^{-8}~\hbox{Hz}}^{-3/2}\myfrac{T_\textrm{obs}}{1~\,\hbox{year}}^{-1/2}\nonumber \\
    &\approx 2\times 10^{-8} [\hbox{s}]\myfrac{f}{10^{-8}~\hbox{Hz}}^{-5/2}
    \myfrac{H_0}{100~ \frac{\hbox{km/s}}{\hbox{Mpc}}}\myfrac{\Omega_\textrm{GW}}{10^{-10}}^{1/2}\myfrac{T_\textrm{obs}}{1~\,\hbox{year}}^{-1/2}
\end{eqnarray}

\begin{figure*}
    \centering
\includegraphics[width=\textwidth]{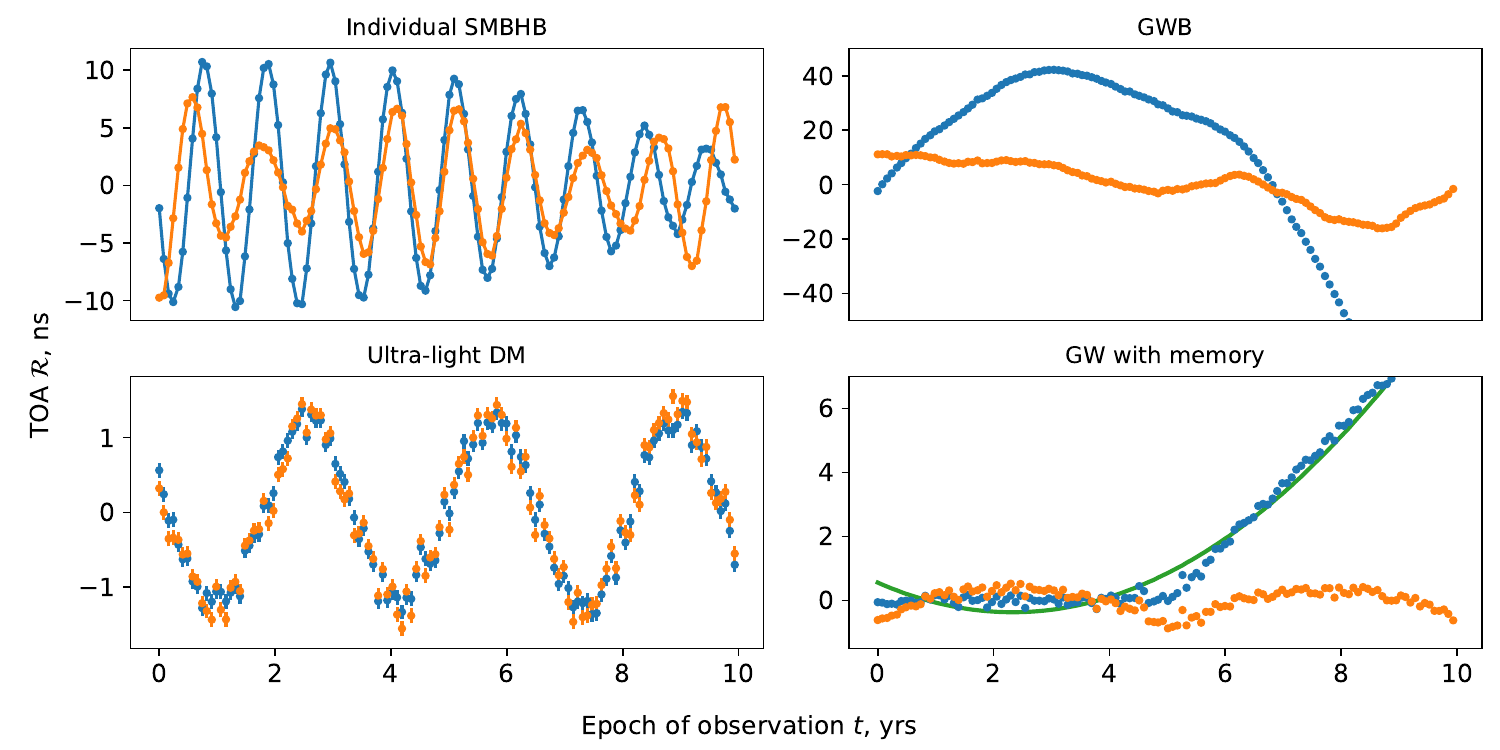}
    \caption{From left to right and from top to bottom: (a) Model residuals from an individual SMBHB with the chirp mass equal to $10^9 M_{\odot}$ and orbital frequency $f=10^{-8}$ Hz located at a distance of $d_L=10$ Mpc; the orange curve shows that the signal contains two frequency harmonics corresponding to short-period Earth and long-period pulsar terms; (b) the model signal from the stochastic GWB with $A=10^{-15}$ and $\alpha=2/3$; (c) the model signal due to the Sachs-Wolfe effect from the ULDM with $m_a=5\times10^{-23}$ eV; (d) the expected residuals from a burst with memory with $h_\mathrm{mem} = 5\times10^{-16}$ and the burst epoch $t_B = 5$~years. For all simulations, the total duration of observations is 10 years. The original signal is shown in blue, and the signal after subtracting the quadratic fitting function is shown in orange. The green line in (iv) shows the fitting function.}

    \label{f:sim_cgw}
\end{figure*}

The signal generated by the GWB is demonstrated in Fig.~\ref{f:sim_cgw}b. Residuals with RMS of tens of ns are potentially measurable by present-level PT experiments. 

For the characteristic amplitude $h_c$ a power-law parametrization is  often used:
\begin{equation}
\label{e:hcAalpha}
    h_c(f)=A\myfrac{f}{f_0}^\alpha\,,
\end{equation}
where the reference frequency is usually set to $f_0=1$~year$^{-1}$. Then the relation of the signal power spectrum in the residuals caused by the stochastic GWB with the amplitude $A$ and the spectral index $\alpha$ is given by (substituting Eq.~(\ref{e:hcAalpha}) into Eq.~(\ref{e:Sfhc})):
\begin{equation}
    S_n(f)=\frac{A^2}{12\pi^2 f_0^{2\alpha}}f^{-\gamma}, \quad \gamma=3-2\alpha.
\end{equation}
Such a model arises both in the classical astrophysical model and in many cosmological scenarios. At least such a parametrization is sufficiently valid approximation for small SNR regimes (see the explanation in the following sections).

\subsection{Pulsar Timing Arrays and stochastic GWB}
\label{s:PTGWB}

Pulsar Timing Arrays (PTAs) are primarly designed to detect stochastic GWB signals and long-period signals from individual sources, such as merging SMBHBs. The sought signal is not transient, but stays in the frequency window of the PT sensitivity with a width $\Delta f\sim 1/ T_\textrm{obs}$ for a substantial amount of time. Thus, there is an accumulation effect: an increase in the total time span of observations $T_\textrm{obs}$ in combination with the ever-increasing sensitivity of the radio telescopes will inevitably increase the SNR and the significance of the detection.

For a PTA observing $N$ identical pulsars over time $T_\textrm{obs}$ with sampling period $\Delta t$ and ToA error $\sigma_n$, the SNR of GWB will grow with amplitude $A$ as: 

$$
\mathrm{SNR}\propto \sqrt{N_\mathrm{psr} (N_\mathrm{psr}+1)} \Delta t^aA^bT_\textrm{obs}^c/\sigma_n^d\,,
$$
where the values $a,b,c,d$ depend on the SNR regime we are working at (see \cite{2013CQGra..30v4015S, 2016MNRAS.458.1267V}). In the weak signal regime (when the spectrum of the GWB lies below the level of instrumental noise over the entire frequency range), $a=-1$, $b=2$, $c=3-2\alpha$, and $d=2$. In the case of intermediate sensitivity, when at least for one frequency bin the spectral power of the signal lies above the noise, the parameters are $a=1/(6-4\alpha)$, $b=1/(3-2\alpha)$, $c=1/2$ and $d=1/(3-2\alpha)$. When deriving these expressions, we used a simple model of the power spectral density of metric fluctuations $S_h(f)\sim f^{2\alpha-1}$, corresponding to the power-law parametrization of $h_c(f)$ (see Eq.~(\ref{e:hcAalpha})).

We would like to emphasize that the equations are derived specifically for the SNR of the HD (not the SNR of uncorrelated stochastic signal common to all pulsars), i.e., only the cross-correlated terms were included in the estimates. When taking into account the auto-correlation terms, the SNR will grow as $\sim\sqrt{N_\mathrm{psr}}$ \cite{2024PhRvD.110f3022B}. The expressions above show that PT is aimed at detecting GWs from the initial long-lasting stages of merging of binaries ("inspiraling"\ stage). In this case, the SNR increases as the data are accumulated. It is opposite to the case of the ground-based GW laser interferometers which are designed to detect transient GW signals from the last stages of evolution of compact binary black holes and neutron stars which sweep the detector sensitivity band for several seconds only.

\begin{figure*}
    \centering
    \includegraphics[width=\linewidth]{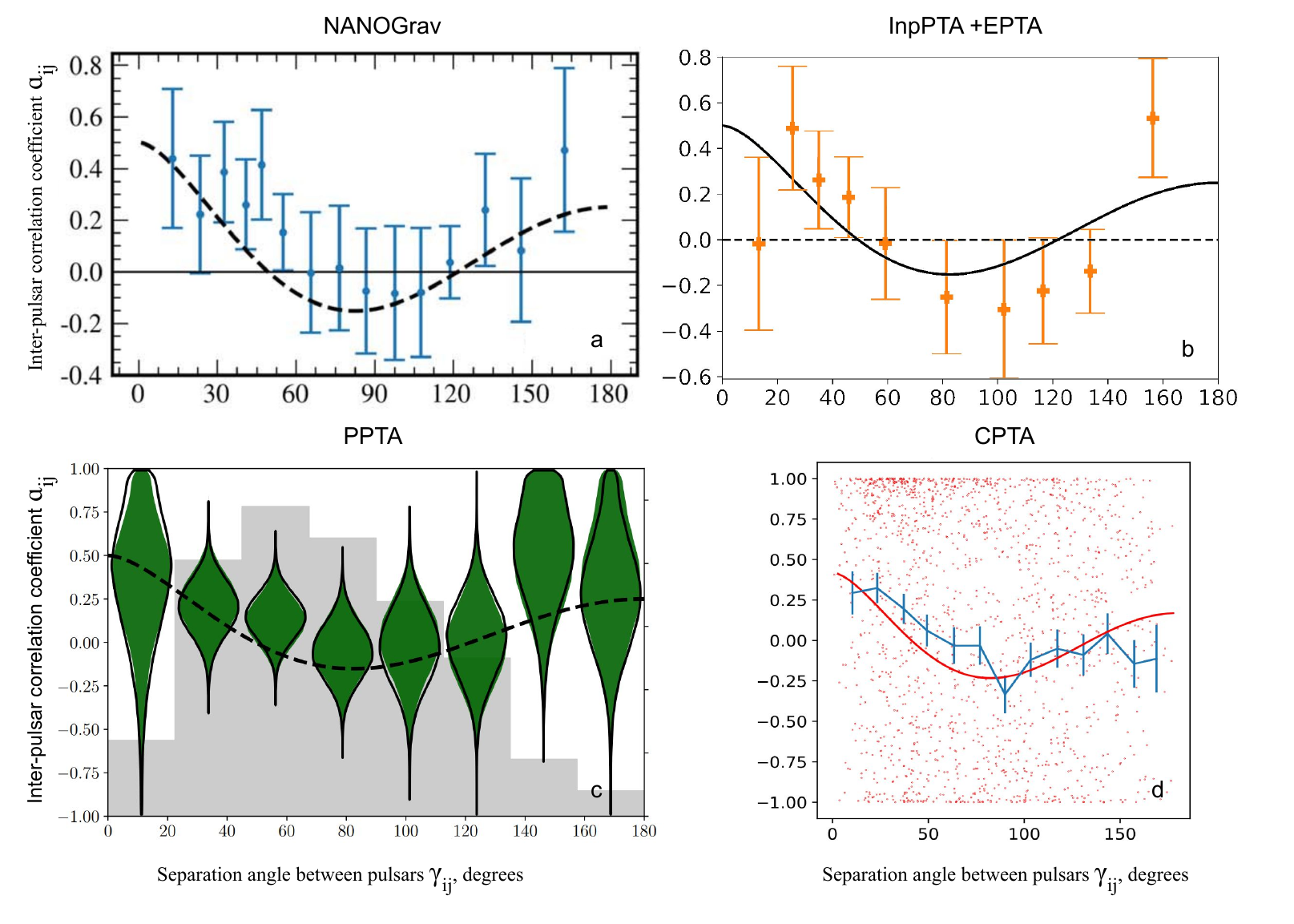}
    \caption{
    Angular correlation $\alpha_{ij}$ averaged for many pulsar pairs as a function of the angular distance between pulsars $\gamma_{ij}$ obtained by different PTAs (left-to-right, top-to-bottom): NANOGrav, EPTA (DR2New), PPTA, CPTA. The theoretically expected curve is shown by dashed (NANOGrav and PPTA), solid (EPTA and CPTA) lines, respectively. For NANOGrav, EPTA, PPTA, the  sought signal is a stochastic process with a spectrum described by a power-law function. For CPTA, a monochromatic signal was searched for in the three lowest frequency windows. The graph shows the resulting angular correlation for one of them, $f=1.5/T_\textrm{obs}$, where $T_\textrm{obs}$ is the total CPTA observation time. Additionally, the correlations of individual pulsar pairs are marked with red dots. From the papers \cite{2023ApJ...951L...8A}, \cite{2023A&A...678A..50E}, \cite{2023ApJ...951L...6R}, \cite{2023RAA....23g5024X}.
    }
    \label{f:corr_pta}
\end{figure*}

In 2023, the major PTA Collaborations, the North American NANOGrav \cite{2023ApJ...951L...8A}, the European  EPTA and Indian InPTA \cite{2023A&A...678A..50E}, the Australian PPTA \cite{2023ApJ...951L...6R}, and the Chinese CPTA \cite{2023RAA....23g5024X}, independently reported the detection of a stochastic signal (red noise) at nHz frequencies, which shows evidence for the HD correlation for different pulsar pairs and is therefore most likely of GW origin. See Table \ref{t:GWB_astro} for the essential characteristics of the aforementioned PTAs: the number of pulsars, the duration of observations, and the observing frequency bands. The analysis was performed within the framework of frequentist and Bayesian statistics \cite{Romano2017}. The significance of the HD angular correlation found ranges from $2\sigma$ to $4.6\sigma$ and is highly dependent on the chosen dataset and processing pipeline. In Fig.~\ref{f:corr_pta} we present the results of the search for angular correlations of the PT residuals independently published by international PTAs. The spectral characteristics of the signal found by different PTAs assuming a theoretically motivated slope of the power spectrum of residuals  $\gamma=13/3$ ($\alpha=-2/3$), are summarized in Table~\ref{t:GWB_astro}. Properties and possible interpretation of the detected signal are discussed in more detail in the next section.

 \begin{table}[]
    \centering
 {\footnotesize    
    \begin{tabular}{cccccc}
    \hline
    Network& Duration  & Number of    & Frequency & Amplitude $A/10^{-15}$ $^\dag$ & Ref.\\
        & (years)  & pulsars $N_\mathrm{psr}$& (MHz) & ($f=10^{-7}-10^{-9}$Hz)\\
        \hline\hline
        NANOGrav & 15 &67 & 327-3000& $2.4^{+0.7}_{-0.6}$ & \cite{2023ApJ...952L..37A}\\
    EPTA/InPTA & 10.3/25 & 25 & 300-4857& $2.5\pm 0.7$ & \cite{2024A&A...685A..94E}\\
    PPTA & 18 & 30 & 692-4032 & $2.04^{+0.25}_{-0.22}$ & \cite{2023ApJ...951L...6R}\\
    CPTA & 3.5 & 57 & 1250 & $2^{+8}_{-80}$ & \cite{2023RAA....23g5024X}\\
    \hline
    \multicolumn{5}{l}{${}^\dag$ For $f_0=1$ yr$^{-1}$, isotropic GWB from coalescing SMBHs $\gamma=13/3$}\\
    \end{tabular}
    \caption{The amplitude of the astrophysical stochastic GW background at $f_0$=1~year$^ {-1}$ with slope $\alpha=-2/3$ from observations of various PTAs. For EPTA, the results of analyzing the latest 10-year data segment (EPTA DR2New) are shown. InPTA data are available for only 10 of the 25 pulsars observed by the EPTA.
    \label{t:GWB_astro}}
        }
\end{table}

\subsection{Signal from individual SMBHBs}

The most expected GW sources in the range of PT sensitivity are SMBHBs, which are assumed to form in the centers of galaxies in the aftermath of the merger process in the model of hierarchical clustering {\cite{2003ApJ...582..559V}. The metric perturbation from such a binary is represented by a monochromatic wave with frequency $f$ twice the orbital one (for circular orbits). The response of a PTA to such a signal includes the Earth and pulsar terms describing the metric perturbations near the Earth at time $t$ and near the pulsar at time $t-\tau_a$, respectively \cite{2012PhRvD..85d4034B}:
\begin{equation}
\mathcal{R}(t) = \sum F^A(\hat{\Omega}) [\mathcal{R}_A(t) - \mathcal{R}_A(t-\tau_a)],
\end{equation}
where $F^A(\hat{\Omega})$ is the antenna pattern of the PTA which 
describes how each of the GW polarizations ($+,\times$) perturbs the residuals depending on the source direction $\hat{\Omega}$. The amplitude of the signal in the residuals is determined by the chirp mass of the binary $\mathcal{M}=(M_1M_2)^{3/5}/(M_1+M_2)^{1/5}$ and the photometric distance to the source $d_L$\footnote{The dimensionless amplitude of the perturbations of the metric from the binary is $h=2\mathcal{M}^{5/3}(\pi f)^{2/3}/d_L$. }:
\begin{equation}
\mathcal{A} \equiv\frac{h_c}{2\pi f}= \frac{\mathcal{M}^{5/3}}{d_L(\pi f)^{1/3}}.
\end{equation}
The full  expression for $\mathcal{R}_A(t)$ depends on the GW polarization angle, the orbital phase, and the orbital inclination angle of the binary (see formula (\ref{e:TOA_GW})) and is given in \cite{2024A&A...690A.118E}. The signal from an individual SMBHB with characteristic parameters is shown in Fig.~\ref{f:sim_cgw}a.

The signal from an individual source will also have a spatial correlation described by the HD curve. Therefore, at low SNRs, the stochastic GWB and the deterministic signal from the SMBHB\footnote{Especially with long orbital periods, which are comparable to the total timespan of the dataset.} in PT data will be strongly covariant, which was confirmed by simulations in \cite{2024A&A...690A.118E}. Indeed, as independently shown in \cite{2023ApJ...951L..50A} and \cite{2024A&A...690A.118E}, the detected signal can be described by a monochromatic GW with frequency $\sim 4$~nHz. Assuming that the binary system is at a distance of $10$ Mpc (slightly less than the characteristic distance to the galaxy clusters in Fornax and Virgo), the chirp mass of the merging binary is $\mathcal{M}\simeq10^9~M_{\odot}$. 

In addition to blind search, in which there is no prior knowledge on binary sky coordinates, there are also targeted searches for GW signals from candidates based using electromagnetic (EM) surveys \cite{2019NewAR..8601525D, 2012AdAst2012E...3D}. This enables increasing the sensitivity locally for specific regions of the sky.

\subsection{Astrophysical GWB}

As shown in \cite{2015MNRAS.451.2417R} using realistic simulations, the detection probability of a stochastic GWB, which is an incoherent sum of signals from the whole population of SMBHBs, is $\sim 5$ times greater than the probability of detecting a deterministic signal from an individual source. Thus, the stochastic GWB is the most promising signal to be probed by PTAs.

The stochastic GWB from the population of incoherent sources of the same nature in the isotropic and homogeneous Universe can be easily calculated analytically \cite{2001astro.ph..8028P}. 
Given a source number density $n(z)$, the contribution to the GWB at frequency $f$ will be determined only by the GW energy emitted by a single source in the logarithmic frequency interval $fdE_{GW}/df$ in the source rest frame at redshift $z$:
\begin{eqnarray}
    &\rho_\textrm{GW}(f)\equiv\Omega_\textrm{GW}\rho_c=\int n(z) dz\int \frac{dE_\textrm{GW}}{df'}\frac{df'}{1+z}=\nonumber\\
    &\int n(z) dz\int \myfrac{dE_\textrm{GW}}{d\log f'} \frac{d\log f}{1+z}
    \label{e:phinney}
\end{eqnarray}
 (the factor $1+z$ in the denominator takes into account the redshift effect  $f=f'/(1+z)$). 

In the case of circular orbits of SMBHBs merging solely due to the GW emission (ignoring possible additional factors responsible for the  orbital momentum loss), the GW energy radiated at the double orbital frequency in the logarithmic frequency interval is $dE_\textrm{GW}/d\log f'=(\pi G  f')^{2/3}\mathcal{M}^{5/3}/3$. 
Substituting this formula into (\ref{e:phinney}), from (\ref{e:Omegagw}) we find that  the amplitude of the signal is determined by the chirp mass of the binary $\mathcal{M}$ and the comoving number density of merging SMBHBs. 
\begin{equation}
h_c^2 (f) = \frac{4G^{5/3}}{3\pi^{1/3}c^2}f^{-4/3}\int \mathcal{M}\int \mathrm{d} z (1+z)^{-1/3}\mathcal{M}^{5/3}\frac{\mathrm{d}^2n}{\mathrm{d}z\mathrm{d}\mathcal{M}},
\end{equation}
where $z$ is the redshift. In this case, the slope of the spectrum  is, $\alpha=-2/3$ $h_c\sim(f/f_\mathrm{ref})^{-2/3}$:

This dependence has a clear physical meaning: for a system in a circular orbit, the dimensionless amplitude of the metric is $h\sim \mathcal{M}^{5/3}\omega^{2/3}/d_L$, and the rate of change of the orbital frequency due to GW radiation is $\dot \omega\sim \omega^{11/3}$. Consequently, in the interval of GW frequencies $\Delta f \sim f$ there will be simultaneously $N=f dN/df=R\omega/\dot\omega$ sources, where $R$ is the merger rate of the binary systems giving the maximum GW amplitude (in this case SMBHBs). Since the sources are evolving independently, the total stochastic signal will have a characteristic amplitude $h_c=\sqrt{N}h\sim f^{-2/3}$. 

In reality, the GW spectrum has a more complex structure, and a number of additional factors must be taken into account for its calculation \cite{2015MNRAS.451.2417R, 2023ApJ...952L..37A}. The non-zero eccentricities of the binary systems will lead to a redistribution of the power spectral density from lower to higher GW frequencies. Similarly, the interaction of the binary components  with the surrounding gas and stars will result in a loss of power at low GW frequencies. Both phenomena will effectively flatten GW spectra. The finite number of merging binaries will lead to the so-called shot-noise of the brightest isolated SMBHBs dominating the overall stochastic signal, which will, on the contrary, effectively increase the slope of the GWB spectrum.

The results obtained by different PTAs (see Fig.~\ref{f:corr_pta} and Table \ref{t:GWB_astro}) are in general agreement with the astrophysical interpretation \cite{2023ApJ...952L..37A, 2024A&A...685A..94E}. Nevertheless, a number of parameters of the astrophysical GWB model are at the borders of their prior distributions. Namely, the spectrum slopes obtained by all PTAs are marginally higher than the predicted $\alpha=2/3$. Moreover, using semi-analytical numerical modeling, such as \textsc{L-Galaxies} \cite{2024A&A...685A..94E, 2023arXiv231206756S}, it was shown that in order to reproduce the found amplitudes of the GWB, it is necessary to significantly increase the mass of the merging components, which can be achieved by increasing the accretion rate onto the black holes. The latter causes an increase in the luminosity function of quasars, which leads to a contradiction with the observations \cite{2024A&A...686A.183I}.

It should be noted that the results presented in this section were obtained under the simplest assumptions about the properties of the GWB: Gaussianity, isotropy, and a power-law shape of the spectrum. Strictly speaking, these conditions are not fulfilled for the astrophysical GWB. Nevertheless, as it was shown in \cite{2023ApJ...959....9B} using realistic models that take into account the finite number of SMBHBs, the methods of PT analysis within the framework of standard assumptions are able to solve the problem, albeit with a rather significant scatter of the obtained parameters of the GWB and signal significance.

\subsection{Cosmological GWB}
\label{sec:cosm_bck}

Cosmological GWB (CGWB) is probably one of the most interesting target from the point of view of PT, as it carries unique information about the physical processes that occurred in the very first epochs of the evolution of the Universe (inflation, reheating phase, phase transitions etc.; see also \cite{2019A&ARv..27....5B,2001PhyU...44R...1G, 2005PhyU...48.1235G,2016arXiv160501615C, 2018CQGra..35p3001C} and references therein). Given that gravitational interaction is extremely weak, GWs leave thermodynamic equilibrium with other components of the Universe immediately after their generation. Qualitatively we can obtain the following estimate:
\begin{equation}
\frac{\Gamma(\mathcal{T})}{H(\mathcal{T})}\sim \frac{\mathcal{T}^3(G\mathcal{T})^2}{\mathcal{T}^2/M_\mathrm{Pl}}=\myfrac{\mathcal{T}}{M_\mathrm{Pl}}^3\,,
\end{equation}
where $M_\mathrm{Pl}$ is the Planck mass, $G=1/M_{\mathrm{Pl}}^2$ is the gravitational constant, $H(\mathcal{T})$ is the Hubble parameter when the  temperature of the Universe is equal to $\mathcal{T}$. $\Gamma(\mathcal{T})=n\sigma v$ gives the probability of gravitational interaction of particles with the interaction cross-section $\sigma\sim G^2\mathcal{T}^2$, velocity $v\sim 1$ and number density $n\sim \mathcal{T}^3$. The above expression suggests that the interaction rate between GWs and the surrounding matter is less than the Hubble parameter for essentially all temperatures $\mathcal{T}<M_{\mathrm{Pl}}$. Thus, GWs propagate freely in the Universe and carry unaltered information about the processes in which they were created. As a consequence, cosmological GWs offer a unique tool for studying extreme physics at ultrahigh energies at the very early epochs of the Universe, which cannot be probed by other means. Among other things, the detection of the CGWB will enable a probe for New Physics beyond the Standard Model, complementing the existing experiments at the Large Hadron Collider at energies up to $\sim 14$~TeV.

Relic GWs (tensor perturbations of the metric) arise directly from the parametric amplification of the primordial quantum fluctuations during the accelerated expansion during the inflationary stage \cite{1982PhLB..115..189R, 1975JETP...40..409G}. Nearly flat or (weakly) red GW spectra is a common property of the conventional slow-roll inflation, so the power spectral density can be parametrized as $P_T(f)\sim f^{n_T}, n_T=-r/8$, where $r=T/S$ is the ratio of tensor to scalar modes of the primordial cosmological perturbations; $r<0.032$ at 95\% confidence level obtained from the analysis of the CMB\cite{2020A&A...641A..10P}. Assuming that the inflationary spectrum with a constant $n_T$ can be extrapolated over many orders of magnitude down to lower frequencies, the expected amplitude of the CGWB is given by \cite{2016PhRvX...6a1035L}\footnote{With Hubble constant equal to 67 km/s/Mpc.}
\begin{equation}
    \Omega_\mathrm{GW}(f)\approx 1.5\times 10^{-16} \myfrac{r}{0.032}\myfrac{f}{f_*}^{n_T},
\end{equation}
where $f_*\approx 7.7\times 10^{-17}$~Hz is the frequency related to the CMB scale of $k=0.05/\hbox{Mpc}$. It is evident that for $n_T\sim 0$ the predicted amplitude of the standard inflationary CGWB is considerably smaller than the value derived from the recent PTA observations. The latter corresponds to a GWB energy density of $\Omega_\mathrm{GW}\sim 10^{-10}$ (\ref{e:sigmaTOA}).

\begin{table}[]
    \centering
   { \footnotesize
    \begin{tabular}{l|l|c}
    \hline
    Model of the CGWB & Derived parameters & Reference \\
    \hline
    \hline
  Relic GWB from inflation  & $\log r=-12.18^{+8.81}_{-7.00}$, $n_T=2.29^{+0.87}_{-1.11}$ & \cite{2024A&A...685A..94E}\\
  &&\\
    & $\log r=-14.06^{+5.82}_{-5.82}$, $n_T=2.61^{+0.85}_{-0.85}$ & \cite{2023ApJ...951L..11A}\\
  &&\\
  \hline
   Cosmic strings   & $\log G\mu=-10.07^{+0.47}_{-0.36} (\mathrm{BOS}),-10.63^{+0.24}_{-0.22} (\mathrm{LRS})$ $^\dag$ & \cite{2024A&A...685A..94E}\\
  & $\log G\mu=-10.15^{+0.16}_{-0.16} (\mathrm{STABLE-C})^{\S}$ & \cite{2023ApJ...951L..11A}\\
   &&\\
   \hline
  MHD turbulence & $\lambda_*\mathcal{H_*}=1, \Omega_*=0.3, T_*=140$~MeV $^\ddag$& \cite{2024A&A...685A..94E}\\
  from QCD phase transition &&\\
  &&\\
  \hline
   Scalar-induced & log$A_\zeta^{10\mathrm{yr}}=-2.9^{+0.42}_{-0.46}, n_s=2.1^{+0.25}_{-0.32}$ &\cite{2024A&A...685A..94E}\\
  2nd-oder GWs&&\\
  ($P_\zeta=A_\zeta^{10\mathrm{yr}}\myfrac{k}{k_{10\mathrm{yr}}}^{n_s-1}$)$^\natural$ &  &\\
  &&\\
  \hline
  Scalar-induced & log$A_\zeta>-1.7$, log$(k_*/\hbox{Mpc}^{-1})>7.6$ $^\S$ &\cite{2024A&A...685A..94E}\\
  2nd-order GWs &&\\
  ($P_\zeta=A_{\zeta}\delta(k-k_*)$) & log$A_\zeta=-0.69^{+0.47}_{-0.47}$, log$(k_*/\hbox{Mpc}^{-1})=-8.9^{+0.60}_{-0.60}$ &\cite{2023ApJ...951L..11A}\\
  &&\\
  \hline
   Phase transitions  & & \\
   in the early Universe &
   $\log\lambda_*\mathcal{H_*}=-0.81^{+0.36}_{-0.36}, \log\alpha_*=0.3^{+0.44}_{-0.44}, \log(T_*/\mathrm{GeV})=-0.76^{+0.36}_{-0.36}$ & \cite{2023ApJ...951L..11A}\\
   &&\\
   \hline
   
    \multicolumn{3}{l}{$^\dag$ Cosmic string tension in BOS model \cite{2015PhRvD..92f3528B}, LRS model \cite{2010JCAP...10..003L}}\\
    \multicolumn{3}{l}{$^\S$ Cosmic string tension in BOS model, when only emission from cusps is taken into account}\\
    \multicolumn{3}{l}{$^\ddag$  $\lambda_*\mathcal{H_*}$  is the ratio of the characteristic length scale of the turbulence, $\lambda_*$, to the comoving Hubble horizon}\\ 
    \multicolumn{3}{l}{ $\mathcal{H_*}^{-1}$ at the QCD epoch with temperature $T_*$ and energy density of the MHD turbulence $\Omega_*$}\\
     \multicolumn{3}{l}{$^\natural$ Power spectrum of the primordial scalar perturbations is normalized to $k_{10\mathrm{yr}}=\frac{2\pi}{10\hbox{yr}}$.}\\
     \multicolumn{3}{l}{{$^\S$} Upper limits at a 95\% confidence level.}\\
    \end{tabular} 
  
    \caption{Parameters of different GWB models under the assumption that the common red noise detected in PTA data has a cosmological origin.
    }
    \label{t:GWB_cosmol}
    }
    
\end{table}

Another process that may contribute to the generation of the background in the early Universe is phase transitions. As the plasma temperature decreases, the Universe undergoes first- and second-order phase transitions, the former being of particular interest in the context of PTAs. For the first-order phase transitions, the average value of the phase of the Higgs field $\langle \phi_T \rangle=0$ changes abruptly to a state with $\langle \phi_T \rangle \neq 0$, which is favored thermodynamically. It is not possible for such a change in the system to occur in the entire space at the same time. The transition is characterized by the formation of bubbles of the new phase and their subsequent expansion at velocities approaching the speed of light (see \cite{2013PhyU...56..747K} for more details on the physics of phase transitions and the early Universe baryogenesis initiated by them). The primary consequence of the expansion of bubbles is the heating of the surrounding plasma, and this process does not result in generation of GWs. Nevertheless, a fraction of the kinetic energy of the scalar field is transferred to the bulk motion of matter and sound waves. By the end of the phase transition, the bubbles merge with additional energy release. The last two processes (bulk flow and bubble merging) result in effective generation of GW radiation due to the non-zero stress tensor $\Pi$. The shape of the power spectral density of the GWB produced by the bubble collision, when the temperature and density of the Universe are equal to $T$ and $\rho_\mathrm{tot}$ respectively, is given by:
\begin{equation}
\Omega_\mathrm{GW}(f)\simeq 1.6\times 10^{-5} \myfrac{100}{g_*(\mathcal{T})}^{1/3}\myfrac{H(t)}{\beta}\myfrac{\Pi}{\rho_\mathrm{tot}}^2\,,
\end{equation}
where $g_*$ is the effective number of relativistic degrees of freedom, $1/\beta$ is the characteristic timescale of the phase transition. For sound waves the energy density of the generated GWB is:
\begin{equation}
\Omega_\mathrm{GW}(f)\simeq 1.6\times 10^{-5} \myfrac{100}{g_*(\mathcal{T})}^{1/3}\myfrac{H(t)}{\beta}\myfrac{\Pi}{\rho_\mathrm{tot}}^2 v_{w} \myfrac{f}{f_0} \myfrac{7}{4+3(f/f_0)^2}^{7/2}\,,
\end{equation}
where $f_0\simeq 2\sqrt{3}(\beta/v_{w})$, $v_{w}$ is the bubble expansion velocity. Both these expressions are interpolations of the results of numerical calculations.
In addition, first-order phase transitions instigate a plethora of further processes, consequently resulting in the generation of GWs within the frequency range that PT is sensitive to. Among those are the MHD turbulence of the primordial plasma \cite{1994PhRvD..49.2837K, 2002PhRvD..66b4030K, 2002PhRvD..66j3505D, 2006PhRvD..74f3521C, 2007PhRvD..76h3002G, 2009JCAP...12..024C} and cosmic strings \cite{1976JPhA....9.1387K, 1995RPPh...58..477H, 2007PhRvL..98k1101S, 2012PhRvD..85l2003S, 2023PhRvD.108l3527Q}, which are topological defects that emit GWs through bursts of cusps, kinks and kink-kink collisions on the loops. 

Another interesting mechanism responsible for the generation of GW radiation relates to the scalar curvature perturbations in the early Universe. The magnitude of such scalar perturbations are associated with the primordial density inhomogeneities which are being measured using CMB temperature maps. The latest data from \textit{Planck} space telescope \cite{2020A&A...641A..10P} suggests that the power spectrum of scalar perturbations $P_\zeta(k)$, measured at scales $k\sim10^{-4}-1$ Mpc$^{-1}$ is nearly flat with the amplitude $A_\zeta=2\times10^{-9}$.

In the linear approximation of Einstein's equations, scalar, vector and tensor modes propagate independently of each other. But in the second order of the perturbation theory scalar modes can launch propagating tensor modes (GWs). In particular, for the power-law spectral density of the primordial fluctuations 
$P_\zeta=A_\zeta^{10\mathrm{yr}}(k/k_{10\mathrm{yr}})^{n_s-1}$ the GW energy is:
\begin{equation}
\Omega_\mathrm{GW}(f=\frac{kc}{2\pi})\simeq (A_\zeta^{10\mathrm{yr}})^2\myfrac{k}{k_{10\mathrm{yr}}}^{2(n_s-1)}\,.
\end{equation}

Therefore, by probing such CGWB, PT enables the estimation of scalar curvature perturbations on much smaller scales than typical CMB scales, with the characteristic wave numbers $k\sim10^{6}-10^{8}$ Mpc$^{-1}$. Probing inhomogeneities on such scales could help us to determine the number of primordial black holes (PBHs) with characteristic masses of the order of several tens of $M_\odot$. Such PBHs are produced from high peaks in the density field when their characteristic size becomes comparable to the size of the cosmological horizon:
\begin{equation}
M(k)\sim 30 M_\odot \myfrac{10.75}{g_*}^{-1/6}\myfrac{3\times 10^{5}\hbox{Mpc}^{-1}}{k}\,,
\end{equation}
$g_*$ is the number of relativistic degrees of freedom.
 
The implications of the signal detected by different PTAs assuming that it is generated by some popular CGWB models, are summarised in Table~\ref{t:GWB_cosmol}. Different CGWB models can explain the observed PTA signal. These models differ in their spectral properties, thereby potentially enabling to distinguish one cosmological model from another (and from astrophysical background). Presently, both the astrophysical and cosmological hypotheses provide equally accurate descriptions of the spectrum of the detected signal \cite{2023ApJ...951L..11A, 2024A&A...685A..94E}. Moreover, based exclusively on the frequency representation of the signal, some of the CGWB models explain the obtained data better than the GWB from SMBHBs \cite{2023ApJ...951L..11A}. However, this requires a non-standard set of model parameters in order to significantly boost the production of GW radiation in the early Universe. It should be noted that these conclusions are likely to be the subject to further revision as the SNR of the detected signal increases and more complete and realistic aspects of the astrophysical GWB are taken into consideration, such as the effects of the eccentricity of SMBHBs and their interaction with the stars and gas in galaxies.

\section{Probing ultra-light scalar field DM using PT and pulsar polarimetry}
\label{s:ULDM}

High-precision observations of MSPs carried out by international collaborations can be used not only to detect GWB and GWs from individual sources, as was discussed in the previous section. In the pioneering paper by V.A. Rubakov and A. Khmelnitsky \cite{2014JCAP...02..019K}, it was shown that the presence of the ultra-light scalar field DM in the Galaxy will induce quasi-monochromatic oscillations in the ToAs of pulses from a pulsar. In addition, the model gives rise to the effect of cosmic birefringence \cite{Carroll:1989vb, Harari:1992ea} resulting in variations of the orientation of the plane of linearly polarized electromagnetic radiation. The constraints on the ultralight scalar DM obtained from PT and pulsar polarimetry are examined in this section of the review.

\subsection{Ultra-light pseudoscalar bosons as DM candidate}

DM continues to be at the cutting edge of fundamental and experimental research in physics, astrophysics and cosmology. \cite{2017PhyU...60....3Z,2018RvMP...90d5002B}. One of the possible candidates for the role of DM are ultra-light, weakly interacting pseudoscalar bosons with masses of $m_a\sim 10^{-22}$\,~eV. 
Independent of the DM problem, axions and axion-like particles (ALPs) have also emerged as a consequence of various extensions to the Standard Model. In particular, ALPs were introduced to solve the CP violation problem of particle physics \cite{2010ARNPS..60..405J}. 
The attractiveness of these particles in cosmology as ultralight DM (ULDM) \cite{1983PhLB..120..127P,1983PhLB..120..133A,1983PhLB..120..137D} is explained by their rich phenomenology. In particular, such bosons with masses $\sim 10^{-20}-10^{-23}$ eV could form so-called "fuzzy"\ DM \cite{2000PhRvL..85.1158H,2017PhRvD..95d3541H}, which solves some of the problems of the cold DM. 

Extremely low masses of ALPs combined with characteristic virial velocities of the order of $v\sim 10^{-3}$  make their de-Broglie wavelength $\lambda_\mathrm{dB}=1/k\sim 1/mv$ ($\geq100$ pc) much larger than the average distance between these particle, i.e. the phase space occupation numbers are huge, 
\beq{e:N}
{\cal N}=\frac{N}{d^3xd^3k}\simeq n\lambda_\mathrm{dB}^3\sim 10^{95}\myfrac{\rho_\mathrm{DM}}{0.4~\hbox{GeV}\,\hbox{cm}^{-3}}\myfrac{m_a}{10^{-22}~\hbox{eV}}^{-4}.
\eeq 
This allows us to treat their ensembles as purely classical fields (see review \cite{2021ARA&A..59..247H}). 

The classical Bose condensate of ALPs in the Galaxy gives rise to a number of significant phenomenological consequences, which may be tested in astrophysical measurements of pulsars. 

\begin{enumerate}
    \item ALP field with boson masses of around ${\cal O}(10^{-22})~\hbox{eV}$ oscillates coherently with typical frequencies defined by the de-Broglie wavelength $\omega=m_a$ (accurate to $\Delta\omega/\omega\sim v^2$): $\phi=A\cos(\omega t+\alpha(t))$ \cite{2021ARA&A..59..247H}. Since the energy-momentum tensor depends quadratically on the field, the time component is $T_{tt}=\rho_\mathrm{DM}=A^2m_a^2/2$, and oscillating spatial components are proportional to the gradient of the field: $\rho_\mathrm{DM}^\textrm{osc}\sim k^2m_a^2=v^2\rho_\mathrm{DM}$. The angular frequency of these oscillations is twice the boson mass $\omega^\mathrm{osc}=2m_a$\footnote{In natural units $\hbar = c=k_B=1$ frequency has units of mass (energy). In CGS, the relationship between frequency of oscillation and the ALP mass is $f\hbox{[Hz]}\simeq 2.4 \times 10^{-8} (m/10^{-22}\hbox{\,eV})$}. Moreover, up to the first order of $v/c$, only the diagonal components are left: $T_{ij}=-\frac{1}{2}m_a^2A^2(\cos{2m_at+2\alpha(t)})=p\delta_{ij}$. The oscillating pressure leads to fluctuations in the gravitational potential, which can be observed by PT and provides independent constraints on the DM density formed by ALPs \cite{2014JCAP...02..019K}.

    \item The non-renormalisable coupling of massive scalar ALPs to photons alter polarization properties of EM waves propagating through ALP field \cite{Carroll:1989vb,Harari:1992ea} (known as effect of birefringence). By probing axion-induced birefringence constraints on the coupling constant between axions and photons $g_{a\gamma}$ were obtained from the analysis of linearly polarized light of active galactic nuclei (AGN) \cite{2019JCAP...02..059I}, from polarimetric observations of protoplanetary disks \cite{2019PhRvL.122s1101F} and from the analysis of the linearly polarized component of the CMB \cite{2019PhRvD.100a5040F,2018arXiv181107873S}.   
    Because of the high degree of linear polarization and extremely stable emission properties, one of the most powerful tools for constraining ALPs is to observe the oscillations in the orientation of the polarization plane of the EM signal from pulsars. For instance, competitive constraints on the coupling constant $g_{a\gamma}$ were obtained with the PPTA data \cite{2019PhRvD.100f3515C, 2022JCAP...06..014C}.

    
    \end{enumerate}

\subsection{Constraints on ultralight scalar DM from  the Sachs-Wolfe effect}
\label{s:uldm_SW}

The first constraints on the energy density of ultralight scalar DM were obtained in \cite{2014JCAP...02..019K}. A summary of the key findings of this work is given below. Coherent pressure oscillations of ultralight scalar DM at twice the mass frequency $2m_a$ lead to quasi-monochromatic variations of gravitational potentials. Indeed, the metric for small scalar perturbations in the covariant Newtonian gauge is expressed as follows:
\beq{e:Nmetr}
ds^2=(1+2\Phi(x,t))dt^2-(1-2\Psi(x,t))dx^2\,.
\eeq
Newton potentials are time-independent $\Psi_0=\Phi_0$ and can be found from the $tt$-components of the Einstein equations: 
$$
\Delta\Psi_0=4\pi GT_{tt}=4\pi G\rho_\mathrm{DM}\,.
$$
so that $\Psi_0\sim G\rho_\mathrm{DM}/k^2$.
Variable gravitational potentials (and similarly for $\Phi(x,t)$) are calculated as a trace of spatial $ij$-components of the Einstein equations:
$$
-6\dfrac{^2\Psi}{t^2}+2\Delta(\Psi-\Phi)=8\pi GT^j\,_j=24\pi G p(x,t)\,,
$$
and oscillate with the same frequency $\omega=2m_a$ as the pressure $p(x,t)$. In the leading order $\Psi_s=0$, and 
\beq{e:Psic}
\Psi_c(x)=\frac{1}{2}\pi G A^2(x)=\frac{\pi G\rho_\mathrm{DM}(x)}{m_a^2}, 
\eeq
and $\Psi_c(x)\sim v^2 \Psi_0(x)$ (as $k^2=m_a^2v^2$). 

EM wave with frequency $f^\textrm{em}_0$ propagating in varying metric (\ref{e:Nmetr}) is time-delayed due to the Sachs-Wolfe effect \cite{1967ApJ...147...73S}. The relationship between the observed frequency $f^\textrm{em}$ at the position of the observer $x$ at time $t$ and $f^\textrm{em}_0$ emitted at $x_0$ of the EM wave propagating in the direction $n_i$, to the leading order (when the Doppler term $n_iv^i\ll 1$ could be  neglected) is:
\begin{equation}
    \frac{\Delta f^\textrm{em} }{f^\textrm{em}}=\frac{f^\textrm{em} -f^\textrm{em} _0}{f^\textrm{em} _0}=
    \Phi(t_0)-\Phi(t)+\int_{t_0}^{t}(\partial_t'\Phi-\partial_t'\Psi)dt'\,,
\end{equation}
where the first two terms are the gravitational redshift of a photon due to the difference of gravitational potentials $\Phi$ at the positions of the observer and the emitter, while the third term represents the integral Sachs-Wolfe effect of an EM wave propagating in the oscillating metric. The integration is performed over the unperturbed trajectory of a photon.
Switching to the full derivatives $\partial_t=d/dt-n_i\partial_i$ one gets:
\begin{equation}
    \frac{f^\textrm{em} -f^\textrm{em} _0}{f^\textrm{em} _0}=\Psi(x,t)-\Psi(x_0,t_0)-\int_{t_0}^{t}n_i\partial_i(\Phi- \Psi)dt'.
\end{equation}

Given that the typical distances to the Galactic sources (pulsars) are usually larger than the characteristic coherence length of the ALP field $D >\lambda_\mathrm{dB} \sim v/m_a$, the expression under the integral is a rapidly oscillating function. As a result, the integral is suppressed by a factor of $k/\omega\sim v\ll 1$. The time-dependent redshift of the signal is only determined by the oscillating part of the potential $\Psi_c$. Therefore, the final expression is given by:
\begin{eqnarray}\label{e:dnunu}
    &\frac{\Delta f^\textrm{em} (t)}{f^\textrm{em} _0} \approx\Psi(x,t)-\Psi(x_0,t_0)=\nonumber\\ &\Psi_c(x)\cos(\omega t+2\delta(x))-\Psi_c(x_0)\cos(\omega (t-D)+2\delta(x_0)) 
\end{eqnarray} 
Note that in this approximation the effect is independent of the gravitational potential $\Phi$. 
At $f =\omega/2\pi=m_a/\pi$ the amplitude of the effect is determined by the local density of the oscillating scalar DM (see Eq.~(\ref{e:Psic})):
\begin{equation}
\label{e:Psic_ma}
    \Psi_c(f )=\frac{\pi G\rho_{\phi}(x)}{m_a^2}=\frac{G\rho_{\phi}(x)}{\pi f ^2}\approx 6.5\times 10^{-18}\myfrac{m_a}{10^{-22}~\hbox{eV}}^{-2}\myfrac{\rho_{\phi}}{0.4~\hbox{GeV\,cm}^{-3}}
\end{equation}

The amplitude of the ALP field is determined by the local DM density, which has a stochastic nature. Consequently, the field amplitude both in the vicinity of the pulsar and Earth is randomly distributed around the mean value $\langle\rho_{\phi}\rangle$: $\rho_\phi(x)=\langle\rho_{\phi}\rangle\kappa^2(x)$, where $\kappa(x)$ follows the Rayleigh distribution.
From (\ref{e:dnunu}), the pulsar timing residuals measured on Earth in $x_e$ for a pulsar located at $x_p=x_0$, can be expressed as:
\begin{equation}
    \mathcal{R}(t)=\int_0^t \frac{\Delta f^\textrm{em} (t') }{f^\textrm{em} _0}dt'=\frac{\Psi_c(x_e)}{\omega}\sin(\omega t+2\delta(x_e))-\frac{\Psi_c(x_p)}{\omega}\sin(\omega (t-D)+2\delta(x_p))\,,
    \label{eq:fdm_sig}
\end{equation}
with a typical induced RMS of
\begin{equation}
    \sigma_\mathrm{ToA}=\sqrt{\langle \mathcal{R}^2(t)\rangle}=\frac{1}{2}\frac{\Psi_c}{2m_a}\approx 0.02 \,\hbox{ns}\, \myfrac{m_a}{10^{-22}~\hbox{eV}}^{-3}\myfrac{\rho_{\phi}}{0.4~ \hbox{GeV\,cm}^{-3}}
\end{equation}
The monochromatic signal from ULDM is shown in Fig.~\ref{f:sim_cgw}c.

In the most general case, the phase of the field on Earth $\delta(x_e)=\delta_e$ (phase of the Earth term) is independent of the phase of the pulsar term $\delta(x_p)=\delta_p$. The same holds for the stochastic amplitudes $\kappa(x_e)=\kappa_e$, $\kappa_(x_p)=\kappa_p$. There are three different configurations:
\begin{enumerate}
\item Uncorrelated case: the distance between the pulsar and the observer is larger than the coherence length of the field $D>\lambda_\mathrm{dB}$; phases $\delta_e$, $\delta_p$ and amplitudes $\kappa_e$, $\kappa_p$ are independent;
    \item Correlated case: all pulsars and the observer fall within the coherence length $D<\lambda_\mathrm{dB}$; phases and amplitudes of the field are the same. Moreover, we assume that the Galacto-centric region is enclosed within the coherence length of the ALP field, so that $\kappa_e=\kappa_p=1$ in the analysis.
    \item  Pulsar-correlated case: the coherence length of the ALP field is larger than the typical Earth-pulsar and pulsar-pulsar distances, but smaller than the Galacto-centric radius. In this case, the DM density estimates obtained from the Galactic rotational curve is averaged over multiple patches, i.e., $\kappa_e=\kappa_p\neq1$.
\end{enumerate}
Depending on the configuration, the ULDM signal has a unique signature of correlation between pulsars. As can be seen from Eq.~\ref{eq:fdm_sig}, if $\rho_\phi(x_e)\gg\rho_\phi(x_p)$, the Earth term dominates over the pulsar term. In this case, the angular correlation between different pulsars is monopolar, i.e. $\alpha_{ij}=1$. Oppositely, if $\rho_\phi(x_e)\ll\rho_\phi(x_p)$ and all pulsars are located in different coherence patches, the monopolar correlation is completely lost. In the most general case, the signal from ALPs exhibits a partially monopolar correlation. As shown in Porayko et al. (2025, PRD submitted), the average correlation coefficient $\alpha_{ij}$ is distributed quasi-uniformly between 0 and 1. This is a distinctive feature of the ALP signal, which allows us to distinguish it from, for example, the GWB with quadrupole correlation.

The concept proposed by A. Khmelnitsky and V.A. Rubakov was developed and validated using real data in \cite{2014PhRvD..90f2008P, 2020JCAP...09..036K, 2023ApJ...951L..11A} (NANOGrav), \cite{2018PhRvD..98j2002P, 2022PhRvR...4a2022X} (PPTA) and using timing of the millisecond gamma-ray pulsars obtained by Fermi-LAT \cite{2023PhRvD.107l1302X}. 
The most stringent constraints were inferred in \cite{2023PhRvL.131q1001S} using the $\sim25$-yr dataset of the EPTA. The density of the ULDM $\rho_{\phi}$ was concluded to be less than several tenths of the local DM density (derived independently from kinematic measurements) for the ALP masses $m_a\sim[10^{-24}~\mathrm{eV}, 10^{-23.3}~\mathrm{eV}]$ (see Fig.~\ref{f:ALMPTA}). 

\begin{figure}
    \centering
    \includegraphics[width=\linewidth]{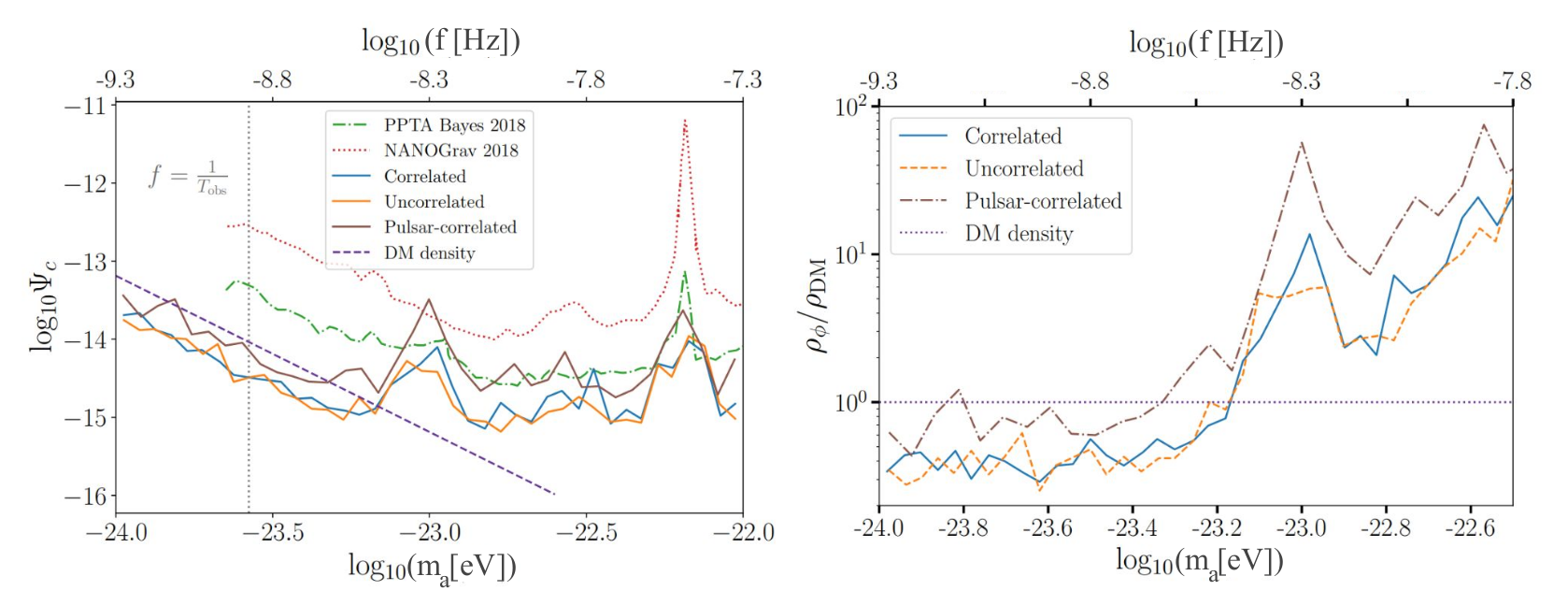}
    \caption{95\% upper limits on the dimensionless amplitude of the ALP field $\Psi_c$ (left panel) and on the ULDM density normalised to the local DM density, $\rho_\phi/\rho_\mathrm{DM}$ (right panel). The bottom horizontal axis shows the ALP mass (eV), while the upper horizontal axis is expressed in terms of the corresponding frequency $f =\omega/2\pi=m_a/\pi\approx 4.8\times 10^{-8}~\hbox{Hz}(m_a/10^{-22}~\hbox{eV})$. The solid lines show the obtained limits for three different PTA configurations (see the text). The dashed-dotted and dotted lines correspond to the previously obtained limits in \cite{2014PhRvD..90f2008P,2018PhRvD..98j2002P}, respectively. The grey vertical dotted line highlights the frequency of $1/T_\textrm{obs}$. The dependency of the amplitude of oscillation on the boson mass (Eq. (\ref{e:Psic_ma}) is shown by the purple dotted line. The figure is adopted from \cite{2023PhRvL.131q1001S}.}
    \label{f:ALMPTA}
\end{figure}

The methodology described above can be generalized to the case of vector ULDM \cite{2020EPJC...80..419N}. The main difference from the case of scalar DM is a non-trivial correlation pattern between ToAs of different pulsars, and on average a four times larger effect than for the scalar DM.

\subsection{Constraints on the ultralight scalar DM from pulsar polarimetry}
\label{s:uldm_birefr}

The non-renormalisable coupling $g_{a\gamma}$ between the scalar $\phi$ and the EM $F_{\mu\nu }$ fields manifests itself in the Lagrangian as follows:
\begin{equation}\label{e:Lagrangian}
    \mathcal{L}=-\frac{1}{4}F_{\mu\nu }F^{\mu\nu }+\frac{g_{a\gamma}}{4} \phi F_{\mu\nu }\tilde{F}^{\mu\nu }+\frac{1}{2}\left(\partial_{\mu}\phi\partial^{\mu}a\phi-m_{a}^{2} \phi^2\right)\,,
\end{equation}
where $\tilde F^{\mu\nu }$ is the EM tensor.
Note that the constant $g_{a\gamma}$ has the dimension of inverse mass. The equations of motion for the Lagrangian (\ref{e:Lagrangian}) are:
\begin{equation}
\label{e:eqmotion}
    \partial^2_t\mathbf{A}-\nabla^2\mathbf{A}=g_{a\gamma}(\partial_t \phi\nabla\times\mathbf{A}+\partial_t\mathbf{A}\times\nabla\phi)\,, \quad\Box \phi+m_a^2\phi=g_{a\gamma}\mathbf{E}\mathbf{B}
\end{equation}
As was discussed above, the amplitude of the scalar field $\phi$ depends on the DM density $\rho_\mathrm{DM}$
\begin{equation}
    A=\frac{\sqrt{2\rho_\mathrm{DM}}}{m_a}\approx 2.5\times 10^{10}~\hbox{GeV}\myfrac{\rho_\mathrm{DM}}{0.4~\hbox{GeV\,cm}^{-3}}^{1/2}
    \myfrac{10^{-22}~\hbox{eV}}{m_a}
\end{equation}
For considered boson masses and characteristic magnetic fields in the interstellar medium 
$B\sim \mathcal{O}(\mu\textrm{G})$ the term $g_{a\gamma}\mathbf{E}\mathbf{B}\ll m_a^2\phi$ in Eq.~(\ref{e:eqmotion}), given typical values of the coupling constant $g_{a\gamma}\lesssim 10^{-12}$GeV$^{-1}$. In other words, one can neglect the back-reaction of the EM field on the ALP field $\phi$. Then the temporal evolution of the field is represented by the coherent oscillations  with frequency $\omega=m_a$:
\begin{equation}
\phi(t,\mathbf{x})=A(\mathbf{x})(\cos(m_at+\delta(\mathbf{x}))
\end{equation}
Assuming that $A(\mathbf{x})$ and $\delta(\mathbf{x})$ changes on a characteristic timescale $T_a$ significantly larger than the period of ALP oscillation (adiabatic approximation),
$$
T_a=\frac{2\pi}{m_a}\approx 4\times 10^7 \hbox{c}\myfrac{10^{-22}\hbox{eV}}{m_a}\,,
$$
and expressing linearly polarized light as a sum of left- and right-handed circularly polarized waves, one gets the dispersive relation for the polarization states:
\cite{2019PhRvL.122s1101F} \begin{equation}
    \omega^2_{\pm} \mp g_{a\gamma}(\partial_t \phi+\widehat{\mathbf{k}}\cdot 
    \nabla\phi)|k|=0\,.
\end{equation}
In a short-wavelength regime $k\gg m_a$ the dispersion relation for the two modes reduces to:
\begin{align}
    \omega_{\pm}\simeq k \pm \frac{1}{2}g_{a \gamma}\left(\partial_{t}\phi+\nabla \phi \cdot\widehat{\mathbf{k}}\right)\,.
\end{align}
Therefore, for plane EM waves the ALP field acts as a birefringent medium. Similarly to the Faraday rotation, for EM field propagating from the emitter at $x_\mathrm{p}$ to the observer at $x_\mathrm{e}$, the plane of polarization is rotated by an angle:
\begin{equation}
  \Delta \theta =\frac{g_{a\gamma}}{2} \int_{t_\mathrm{p}}^{t_{\mathrm{e}}} \frac{d\phi}{dt} dt= \frac{g_{a\gamma}}{2}[\phi(t_\mathrm{e},x_\mathrm{e})-\phi(t_\mathrm{p},x_\mathrm{p})]\equiv\frac{g_{a\gamma}}{2}\Delta \phi. \label{eq:phase}
\end{equation}

However, in contrast to Faraday rotation, the effect from ALPs is independent of the carrier frequency, and is only defined by the difference of the field amplitudes $\phi$ at the pulsar and the observer. 
Such an approximation holds for timescales and characteristic lengths below $\tau_c=1/(m_a v^2)\sim 10^6~T_a$, respectively (see Fig. \ref{f:birefring})
\begin{figure}
    \centering
    \includegraphics[width=0.6\linewidth]{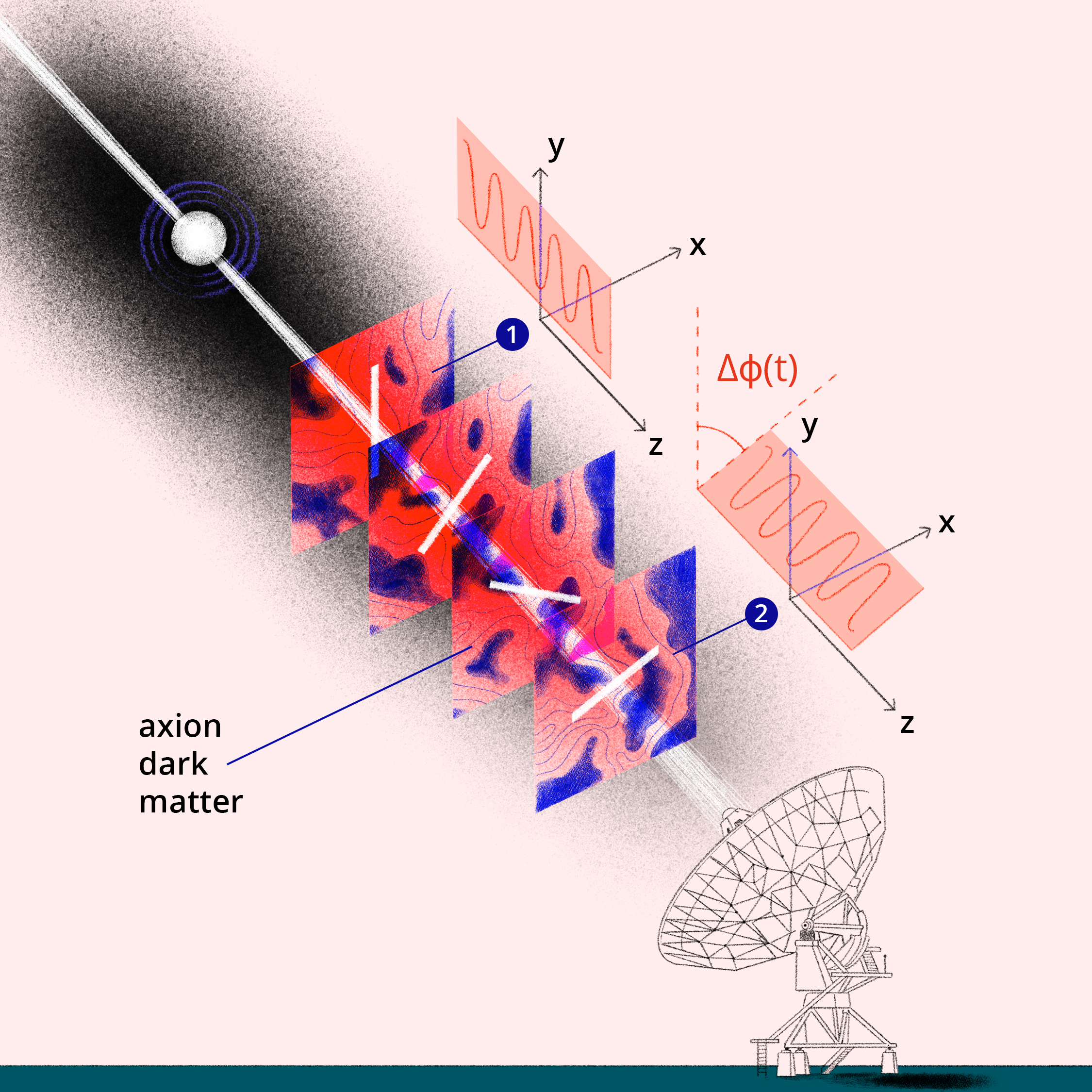}
    \includegraphics[width=0.7\linewidth]{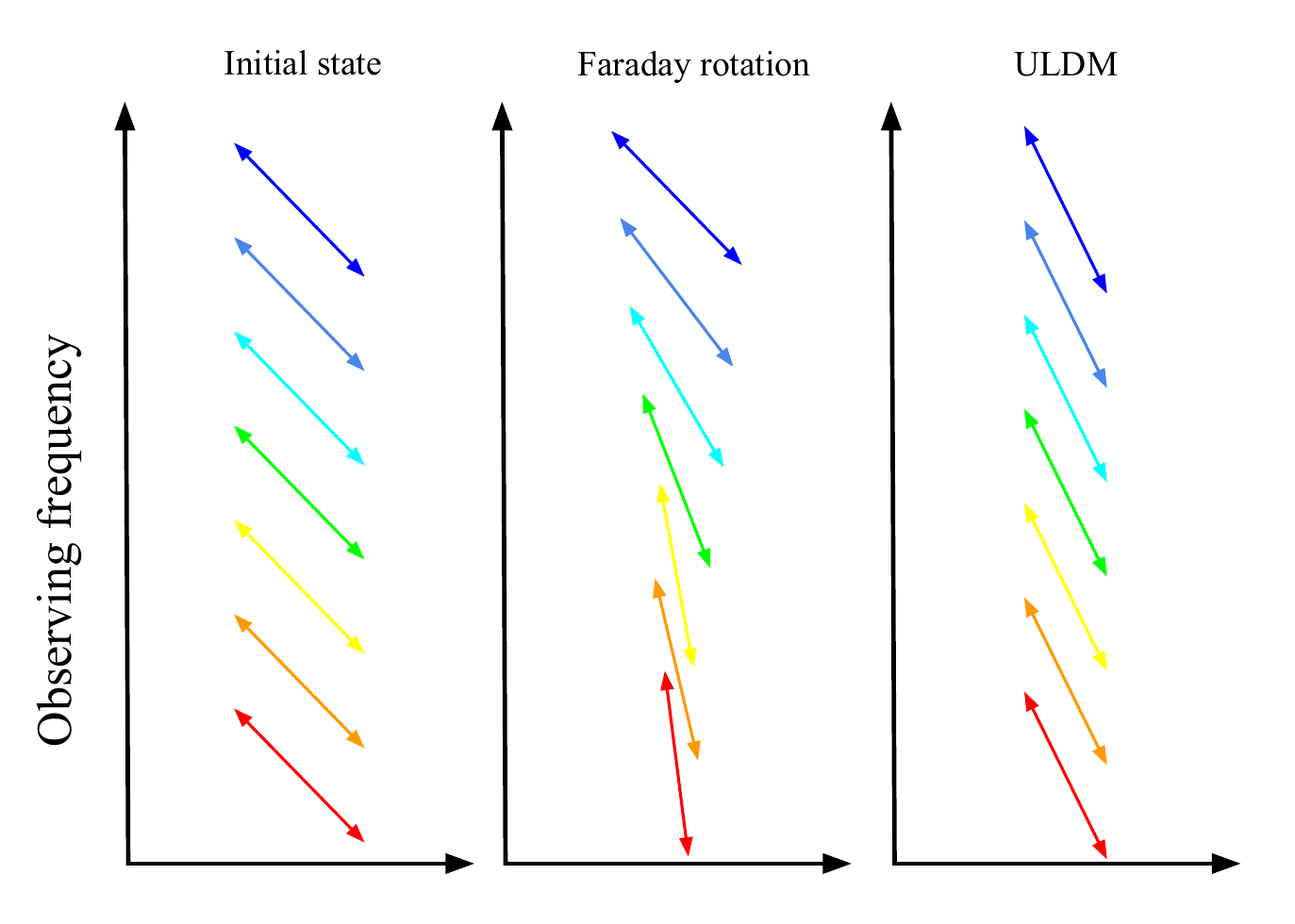}
    \caption{Birefringence of EM wave passing through the ALP field. Upper panel: rotation of the plane of linearly polarized pulsar radiation (adopted from \cite{2022JCAP...06..014C}). Lower panel: change in the polarization properties of the pulsar light due to the Faraday rotation and ULDM.}
    \label{f:birefring}
\end{figure}

\begin{figure}
    \centering
    \includegraphics[width=\linewidth]{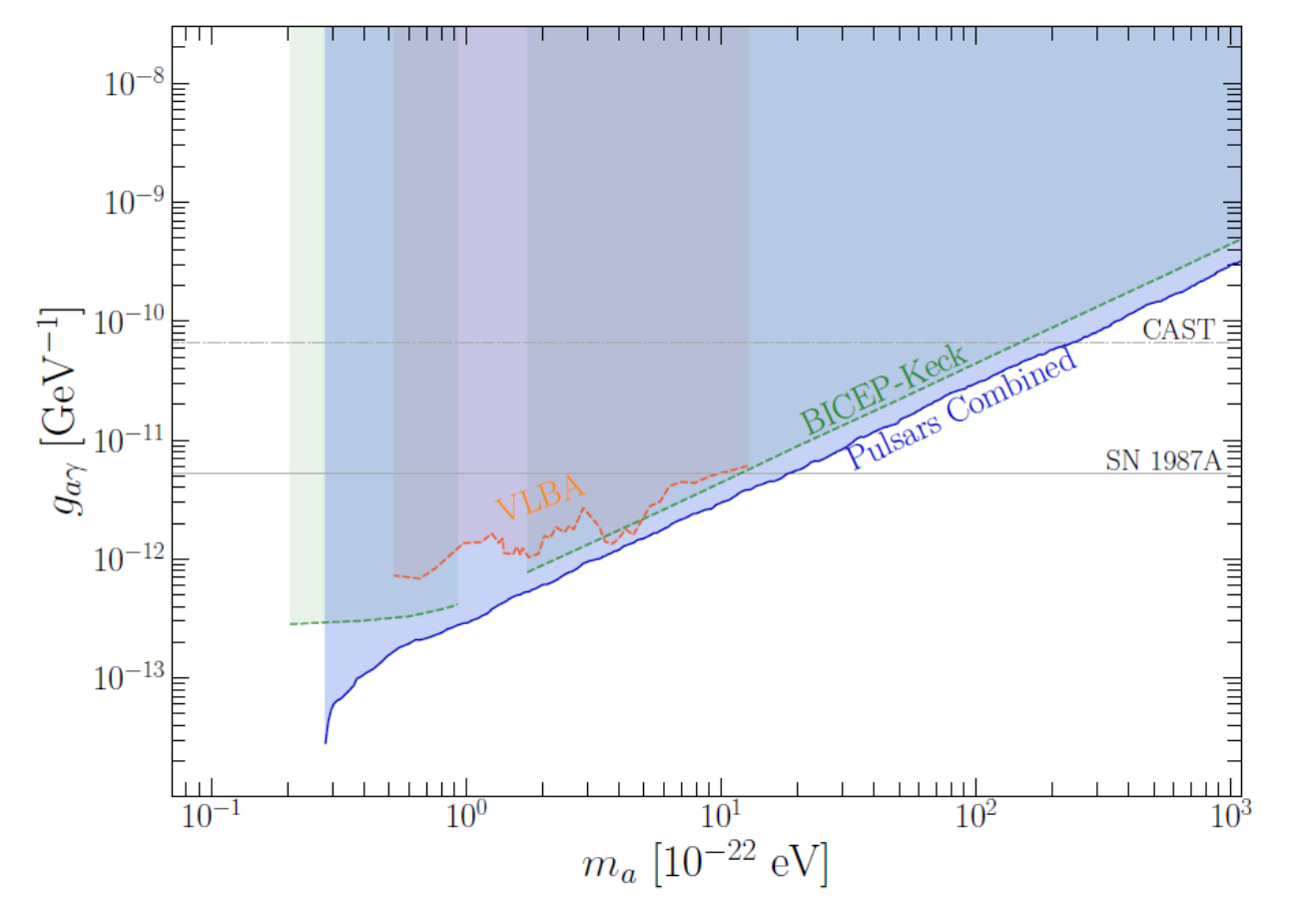}
    \caption{Constraints on the axion-photon coupling $g_{a\gamma}$ as a function of the boson mass obtained from different astrophysical and laboratory probes: helioscope CAST \cite{Cast2017}, axion conversion into gamma photons obtained with SN1987A \cite{Payez:2014xsa}, polarimetry study of AGNs using MOJAVE VLBA \cite{2019JCAP...02..059I} and CMB polarization with BICEP-Keck \cite{BICEPKeck:2021sbt}.}
    \label{f:ALPlim}
\end{figure}

As in the case of the Sachs-Wolfe integral effect, it is necessary to introduce stochastic amplitudes which are used to describe the inhomogeneity of the ULDM density distribution in the Galaxy. The rotation angle of the polarization plane is written as
\begin{align}
\Delta \theta(t) &= \frac{g_{a\gamma}}{\sqrt{2}m_a} \Big[ \sqrt{\rho_\mathrm{e}} \kappa_\mathrm{e} \cos(m_a t + \delta_\mathrm{e})\nonumber\\
&~~~~~~~~~~~~~~- \sqrt{\rho_\mathrm{p}} \kappa_\mathrm{p} \cos(m_a (t - D) + \delta_p) \Big],
\label{eq:full_sig}
\end{align}
where $t$ is the epoch of observation and $D$ is the propagation time from the pulsar to the observer. It is convenient to rewrite this expression in the form:
\begin{equation}
\label{eq:polALP2}
\Delta \theta(t) = \Phi_a \cos(m_a t + \varphi_a),
\end{equation}
where 
\begin{equation}
\label{eq:phi0}
\Phi_a = \frac{g_{a\gamma}}{\sqrt{2}m_a} \left( \rho_\mathrm{e} \kappa_\mathrm{e}^2 + \rho_\mathrm{p} \kappa_\mathrm{p}^2 - 2 \sqrt{\rho_\mathrm{e} \rho_\mathrm{p}} \kappa_\mathrm{e} \kappa_\mathrm{p}\cos\Delta \right)^{1/2},
\end{equation}
with phase $\Delta = m_a D + \delta_\mathrm{e} - \delta_\mathrm{e}$, which, due to large uncertainties in the measured distances to pulsars, can be assumed to be uniformly distributed in the interval $[0,2\pi]$. For typical values $\Delta=\pi$, $\rho_\mathrm{e}=\rho_\mathrm{p}=\rho_\mathrm{DM}$ and $\kappa_\mathrm{e}=\kappa_\mathrm{p}=1$, one gets
\begin{equation}
    \Phi_a=g_{a\gamma}A\approx 
0.025[\hbox{rad}] \myfrac{g_{a\gamma}}{10^{-12}~\hbox{GeV}^{-1}}\myfrac{\rho_\mathrm{DM}}{0.4~ \hbox{GeV\,cm}^{-3}}^{1/2}
    \myfrac{10^{-22}~\hbox{eV}}{m_a}\,.    
\end{equation}
On timescales of a few years, the rotation of the plane of polarization of pulsars typically varies by $1-3^{\circ}$. (see Fig. 2 in \cite{2022JCAP...06..014C}). The ALP birefringence causes periodic oscillations of the polarization plane of pulsar radiation with typical periods of $T_a$. Searching for such periodic components allows us to impose constraints on the coupling constant $g_{a\gamma}$ \cite{2019PhRvD.100f3515C,2020PhRvD.101f3012L,2022JCAP...06..014C}. We would like to emphasize the stochastic nature of amplitudes $\Phi_a$ and phases $\Delta$, which complicates the search for a periodic component in unevenly sampled pulsar observations. Furthermore, the polarimetric measurements can also be biased by a number of other systematic and stochastic effects that are challenging to account for. In particular, the rotation of the plane of polarization of pulsar emission can be attributed to a seasonally varying contribution from the Faraday effect in the Earth's ionosphere. More details on the data analysis methods and newly obtained constraints can be found in Porayko et al. (2025, PRD submitted).

The most recent constraints on the coupling constant $g_{a\gamma}$ as a function of ALP mass $m_a$ are presented in Fig. \ref{f:ALPlim} (the solid blue curve adopted from \cite{2022JCAP...06..014C}), which were derived from polarimetric analyses of 20 PPTA pulsars and QUIJOTE MFI observations of the Crab pulsar used for the PPTA pulsars calibration.

\section{Other applications of pulsar timing}   
\label{s:other}

\subsection{Detection of GW bursts with memory}
The previous sections described the use of PT to detect GWs from SMBHBs at the stage preceding the merger, where the increase in the amplitude and frequency of the signal is relatively slow, i.e., we dealt with quasi-harmonic oscillations. Other types of GW signals also exist. Firstly, at the moment of the merger, there is a short burst with a sharp increase in frequency and amplitude of the signal lasting several tens to hundreds of orbital periods. This type of signals from mergers of stellar mass BHs is detected by LVK laser interferometers.

Secondly, in the so-called "GWs with memory"\ the signal after a GW burst tends to some asymptotic non-zero value \cite{Zeldovich1974, Braginsky1987}. 
This kind of signal arises if during the GW generation there is a permanent change in the value of derivatives of the multipole moments describing the system, which happens if the components of the system before or/and after the GW burst are not gravitationally bound. Initially, the effect was considered for the hyperbolic passages of massive bodies with subsequent escape to infinity and for the supernovae explosion with asymmetric mass ejection. 
The amplitude of the arising GWs at distance $r$ from the source can be roughly estimated as:

\begin{equation}
    \label{eq:bwm1}
   h_{\rm
mem} (r) \sim\frac{r_g}{r}\left(\frac{v}{c}\right)^2,
\end{equation}
where $r_g$ and $v$ are the gravitational radius and velocity corresponding to the non-spherically ejected mass. Obviously, the effect is maximal at a velocity $v=c$, i.e. when the energy is released in the form of photons or GWs.

A significant fraction of the rest mass of merging BHs, up to 10\%, is carried away by asymmetrically radiated GWs \cite{Reisswig2009}, and GW burst with memory (BWM), the so-called Christodoulou effect \cite{Payne1983,Christodoulou1991,Favata2010}, arises naturally.
The amplitude of the BWM for two identical SMBHs with masses $M$ is given by the following expression:

\begin{equation}
\label{eq:bwm2}
h_{\rm mem}(r)=
 5\cdot10^{-16}\left(\frac{M}{10^8~M_{\odot}}\right)\left(\frac{1~\hbox{Gpc}}{r}\right).
\end{equation}

The effect of BWM on pulsar observations was first studied in \cite{Seto2009,Pshirkov2010,vHaasteren2010}. The presence of a constant offset in the metric leads to a linear growth in the residuals starting from the moment of the outburst. One should note that BWM signal mimics the effect of incorrectly determined rotational parameters of the pulsar spin frequency and its first derivative, which leads to a bias in the obtained values after fitting for the timing model.
An example of emerging residuals can be seen in Fig. \ref{f:sim_cgw}d.

As can be seen from this example, the amplitude of the perturbed residuals is quite small even in the case of a merger of relatively close and massive SMBHBs, so it is hardly possible to detect a BWM in observations of a single pulsar. Fortunately, BWMs, like ordinary GWs, cause a correlated response in an array of pulsars, so the observation of a large number of MSPs can significantly increase the sensitivity. In \cite{Pshirkov2010}, we obtained an estimate of the SNR for detecting a BWM signal with the amplitude $h_{\rm mem}$:

\begin{equation}
   \mathrm{SNR}\sim1.5 \left(\frac{h_{\rm
mem}}{10^{-15}}\right)
\left(\frac{N_t}{250}\right)^{\frac{1}{2}}\left(\frac{N_{\mathrm{psr}}}{20}\right)^{\frac{1}{2}}
\left(\frac{T_{\rm obs}}{10~\hbox{years}}\right) \left(\frac{100~~\hbox{ns}}{\sigma_n}\right),
\label{eq:SNR}
\end{equation}   
where $N_{\mathrm{psr}}$ is the total number of observed pulsars, $N_{t}$ is the number of observations of individual pulsars, ${T_{\rm obs}}$ is the total duration of observations, and $\sigma_n$ is the RMS of the ToA residuals. The characteristic values of the parameters are taken for future observations of pulsars with the Square Kilometer Array (SKA) telescope. Assuming the theoretically expected distribution of merging SMBHBs, we can estimate the expected number of detections at the desired SNR level:

\begin{equation}
 N\simeq
10^{-1}\left(\frac{N_t}{250}\right)\left(\frac{N_{\mathrm{psr}}}{20}\right)
\left(\frac{T_{\rm obs}}{10~\hbox{years}}\right)^3
\left(\frac{100~\hbox{ns}}{\sigma_n}\right)^2\left(\frac{3}{\hbox{SNR}}\right)^2\,.
\label{eq:number_of_detection} \end{equation}

BWM are also searched for in the currently available PTA datasets\cite{BWM_Parkes, BWM_NANOGrav}. The best constraints were set by analyzing data from the 12.5-year dataset of NANOGrav. There is a weak indication of a BWM presence, but it may arise from the noise artifacts of individual pulsars in the PTA. More conservative estimates give a 95\%  upper limit on the burst amplitude of $h_{\mathrm{mem}}<3.3\times10^{-14}$.

It should be noted that the similar pattern in the residuals can be caused by the intersection of the LoS to a pulsar by a topological defect, a cosmic string, as a result of the Kaiser-Stebbins effect \cite{Kaiser1984}. An important difference is that the effect now acts only on the individual pulsar, so no correlated signal appears in the entire PTA\footnote{It should be emphasized that in this case we are talking about the signal caused by an individual string on the LoS, not the GWB produced by the string ensemble discussed in \ref{sec:cosm_bck}.}. In this case, the gain is only possible because of the increased number of the potential targets. On the other hand, the amplitude of the effect depends only on the tension of the cosmic string and can be much larger than $\mathcal{O}(10~\hbox{ns})$. In \cite{Pshirkov2010a}, it was shown that pulsar observations allow us to significantly constrain the population of cosmic string loops in the Galaxy for tensions $G\mu/c^2>10^{-14}$.

\subsection{Search for alternative modes of GW polarization}
\label{s:altpol}

In the previous part of the paper we discussed methods of GW search in the framework of GR. In this theory of gravity the GWs are transverse waves and have two states of polarization. The situation can be more complicated in alternative theories of gravity, where additional polarization states can arise, since in metric theories the maximum possible number of states is six \cite{Alves2010}.
Possible additional polarizations will change the response of the GW detectors, which in turn will provide constraints on these alternative theories \cite{Eardley1973}.

The first application of this idea to pulsar timing  was made in \cite{Lee2008}. It was shown that additional polarization states cause significant modifications to the shape of the HD curve (see \ref{sec:h-d}). It  was estimated that the level of sensitivity required to detect conventional GWs would also be sufficient to detect one of the additional states -- the scalar transverse mode, and this level could be achieved by observing 40 pulsars every two weeks for five years at an accuracy level of 100 ns. More detailed calculations were made in \cite{Chamberlin2012,Gair2015}, where it was shown that the sensitivity of PTAs to non-standard modes is much higher than to ordinary ones, especially when correlating signals from pulsars located at a small angular distance from each other on the sky.

The analysis of the observations was used to place constraints on the contribution of non-standard modes: the amplitude of the strongest scalar transverse mode does not exceed $1.4\times10^{-15}$ \cite{NANOGrav_12.5_pol,Wu2022}. The analysis of 15 years of NANOGrav observations, in which the detection of GWs is statistically significant (see the \ref{s:PTGWB} section above), confirms these results \cite{NANOGrav_15.0_pol}.

\subsection{Constraints on graviton mass and GW propagation velocity}

In a similar way, one can obtain constraints on the graviton mass which can be different from zero in many alternative theories \cite{RubakovTinyakov2008,deRham2014}. A non-zero graviton mass would lead to a modification of the HD curve due to two effects. Firstly, the non-zero mass and the corresponding velocity difference between EM waves and GWs reduces the level of correlation \cite{Lee2010}. Secondly, in models with massive gravitons, additional polarization modes are likely to arise and that would affect the shape of the curve as was described earlier in Section \ref{s:altpol} \cite{Liang2021}.
The latest NANOGrav observations limit the graviton mass from above: $m_g<8.2\times10^{-24}~\hbox{eV}/c^2$.

Obviously, questions concerning the graviton mass and the propagation velocity of GWs are closely related. With non-zero graviton mass the propagation velocity of GWs $v_g$ is less than the speed  of light $c$. Simultaneous observations of GW and EM signals from the GW170817 event\cite{GW170817} have placed very strong constraints on the difference between the their propagation velocities, $-3\times10^{-15}<\frac{v_g-c}{c}<7\times10^{-16}$. However, these constraints were derived for $\mathcal{O}(10^2)$~Hz GW frequency. In many models GW velocity may depend on $f$, so that constraints may be much weaker at nHz frequencies to which PTAs are sensitive to \cite{Qin2021}.
Pulsar observations can also limit the superluminal propagation of GWs. In such regimes, the so-called surfing effect will occur and the amplitude of GW-induced residuals will grow resonantly \cite{Baskaran2008}. The absence of the resonant term in the observed residuals allows us to limit the magnitude of the velocity difference $\epsilon\equiv\frac{v_g-c}{c}<10^{-2}$.

In addition, the non-observation of excess noise in the residuals rules out the possibility that massive gravitons may contribute significantly to DM \cite{Pshirkov2008}. This theoretically attractive idea was proposed in \cite{Dubovsky2005}.

A comprehensive review of possible tests of alternative theories of gravity using GW detectors is presented in \cite{Yunes2024}.

\subsection{Study of mass distribution in the Galaxy and search for compact objects}

Non-trivial effects in PT arise when the signal propagates from the source to the observer in non-stationary space-time. If there is a moving gravitating mass near the propagation path, then the spacetime metric is not-stationary. As the result, the non-trivial pattern in the PT residual appears. 
The delay $\Delta t$ in the signal propagation in the vicinity of a gravitating body is known as the Shapiro delay and in the simplest case of a point mass is described as follows \cite{Shapiro1964, Pshirkov2008a}:
\begin{equation}
    \label{eq:shapiro_delay}
    \Delta t=\frac{2GM}{c^3}\ln\left(\frac{4r_\mathrm{lo}r_\mathrm{ls}}{\rho^2}\right),
\end{equation}
where $M$ is the mass of the compact object, $r_\mathrm{lo},~r_\mathrm{ls}$ are the distances from the object to the observer and to the pulsar, respectively, $\rho$ is the impact parameter, the distance of the closest approach between the signal and the gravitating body. The delay consists of two parts: first, the path length increases due to the light bending in the gravitational field of the massive body; second, there is an additional time delay due to the signal propagation in this field.

The Shapiro delay is taken into account in pulsar observations when the gravitating object is a companion of the pulsar in a binary system or belongs to the Solar system. In these cases, it is possible to include the effect into the pulsar timing model with subsequent refinement of the parameters of the object. 
Though, if the effect occurs "on the way"\, it cannot be absorbed by the pulsar timing model, and characteristic features in the pulsar timing residual appear \cite{Larchenkova1995,Pshirkov2008a}.
This effect is similar to the effect of gravitational lensing, and in a similar way PT allows us to study the distribution of gravitating masses in the Galaxy, including its central regions and beyond, where the usual search for gravitational lensing is difficult {\cite{Larchenkova1995,Wex1996, Hosokawa1999,Pshirkov2008a}.

The artifacts in the PT residuals also occur when the gravitating object is extended, and this can be used to search for light mini-halos of DM and to set the corresponding constraints on the mass spectrum of perturbations on small scales \cite{Siegel2007}. The Solar system itself can also act as a detector -- the close passage of a gravitating mass and the resulting shift of the system's barycenter will cause a characteristic response in the residuals of the observed pulsars. This response will exhibit a dipole correlation for different pulsars pairs in the PTA \footnote{Similar effect will be observed when a body passes close to the pulsar, but there will obviously be no correlation in the whole network}. The use of this Doppler effect to search for PBHs of masses about $10^{25}$~g ($\simeq10^{-8}~M_{\odot}$) was proposed in \cite{Seto2007}.

Further potential applications of the PT method were studied in \cite{Baghram2011,Dror2019,Ramani2020,Lee_2021}.
It was shown that the Doppler counterpart of the induced signal makes a major contribution to the potential detection as it is correlated between the pulsars. Future SKA observations of 200 pulsars observed over 20 years at an accuracy level of 50 ns can significantly constrain the abundance of PBHs and mini-halos over a wide mass range of $10^{-8}-10^2~M_{\odot}$, which is larger than that probed by the gravitational microlensing method. The  analysis  \cite{Lee_2021} was applied to 15 years of NANOGrav observations \cite{2023ApJ...951L...8A}. The obtained constraints on the PBHs abundance and their contribution to the total DM density are so far much weaker than theoretically possible estimates, $f\equiv\frac{\Omega_{\mathrm{PBH}}}{\Omega_{\mathrm{DM}}}\sim\mathcal{O}(10^2)\gg 1$.

The weakness of the obtained constraints is related both to much more modest capabilities of NANOGrav compared to the future parameters of the SKA PTA, and to the idealization of the noise characteristics of pulsars in the theoretical estimates. It was assumed that the residuals behave like white noise, while many real pulsars have a significant contribution from the intrinsic "red"\ noise (arising due to the rotation instability of pulsars), which substantially reduces the sensitivity of the PTA to the considered effects.

\section{Perspectives of the PT}

\label{s:conclusion}

At the moment, direct detection of GWs has only been achieved at high frequencies by the LIGO/Virgo/KAGRA ground-based gravitational interferometers. PT uniquely complements these measurements in the nHz range. The recent evidence suggests that a detection of GWs at a high level of significance awaits us in the near future. International pulsar collaborations are undertaking the necessary steps to reach this objective as soon as possible. Further progress in the field of PT is related to three key aspects.

\paragraph{Increasing the significance of detection.}

Combining the data sets of individual regional PTAs within the framework of the IPTA is the most natural step of increasing the SNR of the detected signal. The combination will inevitably lead to both a more complete and regular coverage of the celestial sphere by pulsars and an increase in their total number, which will enable a better measurement of the HD curve. It is expected that as a result of data combination, which is a very labor-intensive process requiring huge human resources, the sensitivity of the final PTA will be doubled and the GW signal will be reliably detected with a significance higher than 5$\sigma$.

Increase in the detection significance can also be achieved by extending the total observation time $T_\textrm{obs}$. However, it is worth noting that the increase in the SNR slows down from $T_\textrm{obs}^{13/3}$ to $T_\textrm{obs}^{1/2}$, when the transition from the weak signal mode to the detection mode takes place. Moreover, as the timespan of observations increases, the contribution of the stochastic processes such as the pulsar's intrinsic noise and the turbulence of the interstellar medium, which both complicate the search for the GW signal, becomes significant. An alternative method to increase the SNR is to use more sensitive radio telescopes for regular monitoring of pulsars, which will improve the timing accuracy and, consequently, reduce the contribution of the instrumental white noise. For example, PTA based on only 4 -- 5 years of data collected with the MeerKAT \cite{2020PASA...37...28B} and FAST \cite{2013MS&E...44a2022N} radio telescopes have sensitivities comparable to those achieved by historical PTA collaborations over the time span of decades. SKA \cite{2015aska.confE..37J} and DSA-2000 (Deep Synoptic Array \cite{2019BAAS...51g.255H}), which are expected to be operational in the next decade, plan to launch extensive pulsar programs that will unprecedentedly increase the sensitivity of future PTAs to various kinds of astrophysical signals.

\paragraph{Optimization of pulsar networks and the algorithms used.}

One of the main challenges in this field of astrophysics is a high computational complexity and the challenge of Big Data. In addition to big data volumes, one needs to solve a multi-parameter optimization problem when searching for GW signal (at present, the number of free parameters of the problem reaches $\sim$300). Therefore, studying various methods of optimization and increasing the speed of computation is one of the prior directions in the PT field. In particular, advanced Markov chain Monte Carlo schemes have been developed\cite{2024arXiv241011944L}, and machine learning methods have been introduced into the PT data analysis\cite{2024PhRvL.133a1402S} schemes. The launch of DSA-2000 and SKA will further increase the volume of processed data due to newly discovered pulsars, which once again emphasizes the relevance of the optimization problem.

\paragraph{Identifying the nature of the signal.}

To date, it is impossible to determine conclusively the nature of the detected correlated stochastic signal in the residuals. We may be dealing with the cosmological GWB generated in the early Universe, or with astrophysical GWs from the entire population of SMBHBs, or from a single bright source. There is also a possibility that the detected signal is a combination of astrophysical and cosmological backgrounds. In \cite{2023ApJ...951L..11A, 2024A&A...685A..94E} attempts were made to identify the signal based solely on its spectral characteristics. But as was shown in the previous sections, the shapes of the spectra of the cosmological and astrophysical backgrounds can vary greatly depending on the underlying assumptions. In particular, the power spectral density of the astrophysical background $S_h(f)$, when taking into account the eccentricities of merging binaries and the finiteness of the number of sources, strictly speaking, is not described by the power law. Thus, considering only spectral characteristics, two types of the GWBs are indistinguishable, and the problem of signal identification becomes degenerate.

However, the degeneracy can be removed by including additional observables in the analysis and using non-standard statistical characteristics of the signal (see, e.g., \cite{Romano2017}). One such characteristic is the spatial anisotropy of the GW signal. The astrophysical GW background, which is an incoherent sum from individual SMBHBs, is expected to have a high degree of anisotropy, especially at high frequencies, where individual sources will dominate over the mean background. As shown in \cite{2024ApJ...965..164G}, some astrophysical models predict a significant level of anisotropy, already exceeding the current upper estimates at $C_1/C_0\leq 0.2$ derived in \cite{2023ApJ...956L...3A}. The sensitivity to anisotropy of the GWB will continue to increase with an addition of new pulsars to a PTA and with decreasing standard RMS $\sigma_n$ of the residuals, as $\sqrt{N_\textrm{psr}/\sigma_n}$ \cite{2024arXiv240714460D, 2022ApJ...940..173P}. The lack of anisotropy in the data obtained with the latest pulsar observational programs will enable us to strongly constrain the parameter space of the merging SMBHBs models or exclude the astrophysical nature of the signal. However, it is worth noting that a number of effects, such as the cosmological variance of the HD curve, which has similar observational manifestations, can strongly weaken our sensitivity to the anisotropy probes \cite{2024arXiv240807741K}.

\section*{Acknowledgements}
The work of KP and MP was supported by grant 075-15-2024-541 of the Ministry of Education and Science of the Russian Federation under the program for financing major scientific projects of the national project "Science". The work of NP was supported by Deutsche
Forschungsgemeinschaft (DFG, German Research Society) -- project number PO 2758/1--1, under the Walter-Benjamin Fellowship.

\bibliographystyle{UFN}
\bibliography{pta}

\end{document}